\documentclass[
    aps,
    physrev,
    twocolumn,
    unsortedaddress,
    superscriptaddress,
    showkeys,
    nofootinbib,
    10pt
]{revtex4-2}

\usepackage{graphicx}
\usepackage{dcolumn}
\usepackage{bm}
\usepackage[dvipsnames]{xcolor}
\usepackage[colorlinks=true,urlcolor=Blue,linkcolor=NavyBlue,citecolor=NavyBlue]{hyperref}
\usepackage{amsfonts}
\usepackage{mathrsfs}
\usepackage{empheq}
\usepackage{amsmath}
\usepackage{amssymb}
\usepackage{cancel}

\graphicspath{{./}{Figures/}}

\begin{document}
\title{Direct detection of right-handed fermionic dark matter: electron-recoil measurements in Xe atoms}
\author{Santiago Collazo} 
\email[]{Contact author: scollazo@fcaglp.unlp.edu.ar}
\affiliation{Instituto de Astrofísica de La Plata (CONICET-UNLP), Paseo del Bosque S/N, Argentina}
\affiliation{Facultad de Ciencias Astronómicas y Geofísicas de La Plata (UNLP), Paseo del Bosque S/N, Argentina}

\author{Carlos R. Argüelles}
\email[]{carguelles@fcaglp.unlp.edu.ar}
\affiliation{Instituto de Astrofísica de La Plata (CONICET-UNLP), Paseo del Bosque S/N, Argentina}
\affiliation{Facultad de Ciencias Astronómicas y Geofísicas de La Plata (UNLP), Paseo del Bosque S/N, Argentina}
\affiliation{ICRANet, Piazza della Repubblica 10, I-65122 Pescara, Italy}

\author{Soroush Shakeri}
\email[]{s.shakeri@iut.ac.ir}
\affiliation{Department of Physics, Isfahan University of Technology, Isfahan 84156-83111, Iran}

\author{Osvaldo Civitarese}
\email[]{osvaldo.civitarese@fisica.unlp.edu.ar}
\affiliation{Departamento de Fisica (UNLP) and Instituto de Física de La Plata (CONICET-UNLP), Diagonal 113 entre 63 y 64, Argentina}

\date{\today}

\begin{abstract}
    We present theoretical estimations of the event rates for the interaction between a right-handed fermionic dark matter with an electron bound to a xenon atom. Motivated by recent astrophysical and cosmological constraints on a fermionic dark matter candidate in the sub-MeV mass range, we study such an interaction through an effective electromagnetic channel where the output consists of standard-model particles. This mechanism allows right-handed neutral dark matter fermions to couple with standard-model particles with the interchange of $W^{\pm}$ or $Z$ bosons, without the need to add extra fields. We generalize and extend previous results on this type of channels, by including for the ionization form factor arising from the interaction of the fermionic candidate with bound electrons. In addition, we compute the sensitivity curves for different direct detection experiments in such a challenging light particle mass range.
\end{abstract}


\maketitle

\section{Introduction}

Dark matter (DM), an invisible component of matter whose gravitational effects are necessary to reconcile cosmological theory with observations, is today the most widely accepted paradigm in astronomy. Dark matter particles are expected to be electrically neutral and not to emit electromagnetic radiation. Understanding the nature and distribution of dark matter is crucial for explaining structure formation in the Universe and galactic dynamics. Cosmological observations have determined that dark matter comprises approximately five times more mass than all baryonic matter combined \cite{PlanckCollaboration2020}.

Given the observational evidence for dark matter, identifying the particle constituent has become central to particle physics. One of the most extensively studied approaches is that proposed by Lee and Weinberg \cite{Lee1977}, who established a lower bound of approximately $m \sim 2$ GeV$/c^{2}$ for a neutral heavy lepton, now known as the Lee-Weinberg limit. This mass scale traditionally defines WIMP particles (Weakly Interacting Massive Particles) in the literature, although the term encompasses a broader range of masses and interaction strengths, including weakly and sub-weakly interacting regimes \cite{Roszkowski2018}.

WIMP dark matter with masses exceeding $\sim 10$ GeV$/c^{2}$ remains a viable candidate for $\Lambda$CDM cosmology. Although this model successfully describes low-redshift large-scale structure (LSS) \cite{Troster2020} and the cosmic microwave background (CMB) anisotropy spectrum \cite{PlanckCollaboration2020}, significant tensions emerge at galactic scales (see \cite{Perivolaropoulos2022} for a comprehensive review). Notable discrepancies include the missing-satellite problem \cite{Kauffmann1993, Klypin1999, Moore1999a}, the too-big-to-fail problem \cite{Tikhonov2009, Boylan-Kolchin2011, Ferrero2012, Boylan-Kolchin2012, Tollerud2014, Garrison-Kimmel2014, Papastergis2015, Kaplinghat2019}, and the core-cusp problem \cite{Moore1994, Flores1994, Salucci2001, deBlok2010}.

One promising candidate for resolving these discrepancies is a fermionic dark matter particle with mass $\mathcal{O}$(keV$/c^{2}$). Such particles, commonly termed sterile neutrinos in the literature, do not interact electromagnetically or via standard model interactions (see \cite{Adhikari2017} for a comprehensive review of keV-scale sterile neutrino astroparticle physics). This class of particles constitutes warm dark matter (WDM) in cosmological terminology \cite{Melott1985}. WDM is characterized by sufficiently high free-streaming velocities to suppress the growth of small-scale primordial density fluctuations. This suppression naturally erases the formation of sub-galactic dark matter substructures, potentially resolving small-scale cosmological anomalies while preserving the successful predictions of $\Lambda$CDM at large scales \citep[e.g.][]{deBlok2001, Boyarsky2009, Oh2011, Lovell2012, Menci2012, Agnello2012, Papastergis2015, Lovell2017}. Although extensive literature addresses WDM with particle masses of a few keV for small-scale structure problems, larger fermionic dark matter masses remain consistent with cosmological constraints \cite{Viel2006, Seljak2006, Viel2008, Polisensky2011, Viel2013, Kennedy2014}.

Motivated by recent astrophysical results on fermionic dark matter halos \citep[see][for a review]{Arguelles2023b} and by the continued absence of positive detections in WIMP-search experiments \cite{PandaXCollaboration2021, XenonCollaboration2022, LZCollaboration2024}, we investigate a neutral fermion candidate with mass $\mathcal{O}(10$--$200$ keV$/c^{2}$). Unlike traditional WDM scenarios, we consider this as a distinct light dark matter (LDM) particle. Assuming a spherical dark matter halo composed of such fermions exists in general relativistic equilibrium, we can study the physical implications of this self-gravitating fermionic system under hydrodynamic and thermodynamic equilibrium. Halos exhibiting dense core--dilute halo structure \cite{Arguelles2018} successfully reproduce astrophysical observables spanning milliparsec to tens of kiloparsec scales. These include the shadow-like image and orbital dynamics around Sagittarius A* \cite{BecerraVergara2020, BecerraVergara2021, Pelle2024}, accretion physics onto compact objects \cite{Millauro2024}, Milky Way rotation curves \cite{Arguelles2018, Arguelles2019, Arguelles2023a}, and stellar stream dynamics \cite{Mestre2024, Collazo2025}.

Direct detection experiments provide a powerful means to probe the microscopic nature of dark matter. Numerous experimental collaborations are actively searching for dark matter particles \cite{CoGeNTCollaboration2013, DAMICCollaboration2015, DarkSideCollaboration2015, DARWINCollaboration2016, COSINECollaboration2018, SENSEICollaboration2018, TheLZCollaboration2019, CRESSTCollaboration2019, Rau2020, PandaXCollaboration2021, DAMALIBRACollaboration2022, XenonCollaboration2024}. This work focuses on electron recoil signatures \cite{Essig2012}. The low LDM mass scale requires sensitivity to energy transfers smaller than those accessible in nuclear recoil experiments. Electron recoil experiments achieve sufficient sensitivity to detect ionization signals from dark matter interactions with atomic electrons. Promising target materials include semiconductors \cite{Essig2012, Graham2012, Lee2015, Essig2016, Hochberg2017}, superconductors \cite{Hochberg2016a, Hochberg2016b, Hochberg2016c, Hochberg2019, Kim2022, Hochberg2023}, and noble liquids \cite{Essig2012, Lee2015, Shakeri2020, Catena2020, Dror2021, PandaXCollaboration2022, Li2022, DarkSideCollaboration2023}. For a comprehensive overview of electron recoil experiments, see \cite{Ge2022}.

We specifically focus on xenon and examine absorption processes, in which the incident dark matter particle deposits both its rest mass energy and kinetic energy into bound atomic electrons. This contrasts with scattering, where only kinetic energy is transferred. 

This work is organized as follows. Section \ref{sec:physical_context_and_formulae} presents the physical framework, the interaction model, the event rate phenomenology, and the differential cross section. Section \ref{sec:results} contains the results, and Section \ref{sec:conclusions} provides the conclusions.

\section{Physical context and formulae}\label{sec:physical_context_and_formulae}

We investigate the flux of right-handed neutral dark matter fermions (RHDMF) incident upon a liquid xenon detector. Xenon is an ideal target material due to its chemical inertness and negligible radiative decay probability. The physical process involves an RHDMF interacting with an electron bound to a xenon atom. This interaction cannot be treated as a simple elastic or inelastic collision since the electron is bound to the quantum potential of the xenon nucleus (treated as static). The nuclear potential significantly affects the scattering dynamics. Following the collision, which transfers energy and momentum to the bound electron, the process produces a left-handed standard model (SM) neutrino and an ionized electron as free particles. This yields a two-particle initial and final state described by the reaction:

\begin{equation}
    \chi(E_{\chi}, \vec{p}_{\chi}) + \mathrm{e}^{-}(E_{k}, \vec{k}) \rightarrow \nu(E_{p'}, \vec{p'}) + \mathrm{e}^{-}(E_{k'}, \vec{k'}).
\end{equation}

Except for the outgoing SM neutrino, the speeds involved are non-relativistic. The bound electron moves at non-relativistic velocities, typical incident DM speeds are $\sim 10^{-3}c$, and the outgoing ionized electron remains non-relativistic. We thus parameterize energies as $E_{\chi} = m_{\chi} + \frac{1}{2}m_{\chi}v^{2}$, $E_{k} = m_{e} + E_{B}^{nl}$, $E_{k'} = m_{e} + E_{R}$, and $E_{p'} = p'$\footnote{We assume the convention $c = \hslash = 1$ along the work.}, assuming massless neutrinos. The binding energy of the electron in the $(n, l)$ shell is $E_{B}^{nl}$ \citep[tabulated for xenon in][]{Bunge1993}. The recoil energy $E_{R} = k'^{2}/2m_{e}$ represents the kinetic energy of the unbound ionized electron.

The momentum transfer vector $\vec{q} = \vec{p}_{\chi} - \vec{p'} = \vec{k'} - \vec{k}$ quantifies momentum transferred to the bound electron by the incoming right-handed fermion. With DM momentum $\vec{p}_{\chi} = m_{\chi}\vec{v}$, referencing the velocity $\vec{v}$ to the detector frame, and the outgoing neutrino momentum $p' = |m_{\chi}\vec{v} - \vec{q}|$, the energy conservation equation yields:

\begin{align}
    m_{\chi} + m_{e} + \frac{m_{\chi}v^{2}}{2} + E_{B}^{nl} =&\ m_{e} + |m_{\chi}\vec{v} - \vec{q}| + E_{R},\notag \\
    m_{\chi} + E_{B}^{nl} - E_{R} =& -\frac{m_{\chi}v^{2}}{2} \notag \\ 
    &+ \sqrt{q^{2} + m_{\chi}^{2}v^{2} - 2m_{\chi}vq\mathrm{cos}(\theta_{qv})}.
    \label{eq:energy_conservation}
\end{align}
Following the approximation of \cite{Dror2021} that $m_{\chi}v \ll q$, since the momentum of the DM particle is much less than the momentum of the outgoing SM neutrino, we can neglect terms of order $v$ and higher in Eq. (\ref{eq:energy_conservation}) to obtain:

\begin{equation}
    q = m_{\chi} + E_{B}^{nl} - E_{R}.
    \label{eq:energy_conservation_approximation}
\end{equation}

Recent fermionic dark matter halo astrophysics \cite{Arguelles2018, Arguelles2022, Arguelles2023a, Arguelles2023b, Arguelles2023c, Millauro2024} motivates studying RHDMF in the mass range $m_{\chi} \in [25, 200]$ keV. The binding energies are shell-dependent. Equation (\ref{eq:energy_conservation_approximation}) thus provides a linear relationship between recoil energy $E_{R}$ and momentum transfer $q$, parameterized solely by the DM mass $m_{\chi}$. This simplifies the calculation of the differential event rate discussed in the following section and in Appendix \ref{ap:event_rates}.

\subsection{Interaction model}\label{sec:interaction_model}

Direct detection of dark matter requires an interaction mechanism between the DM candidate and detector material. We employ the model introduced in \cite{Shakeri2020}, based on the weak-coupling infrared fixed point of a four-fermion interaction of the Nambu-Jona-Lasinio (NJL) type \cite{Nambu1961}. In the infrared regime, such quadrilinear operators induce one-particle-irreducible interacting vertices between left-handed and right-handed fermions. Notably, this framework permits couplings of right-handed currents with the standard model $W^{\pm}$ and $Z^{0}$ bosons \cite{Suhonen1998, Xue1999, Xue2000}.

\begin{figure}
    \centering
    \includegraphics[width=0.5\columnwidth]{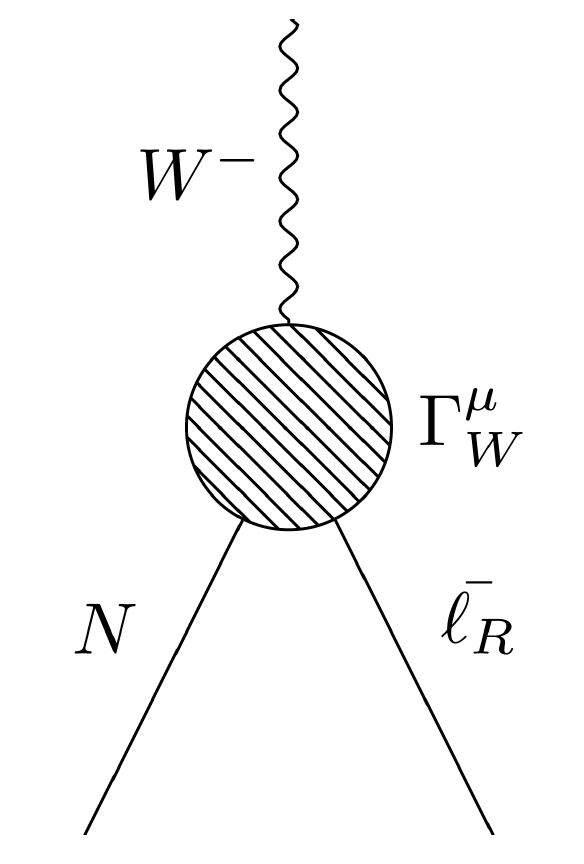}
    \caption{Effective coupling vertex corresponding to the coupling of the neutral RHDMF and right-handed electrons with the $W$ boson.}
    \label{fig:feynman}
\end{figure}

Thus, we consider the coupling between an RHDMF ($N_{R}$) and a right-handed charged lepton with the $W$ boson in the infrared regime. The interaction Lagrangian reads \cite{Xue2003, Shakeri2020}:

\begin{equation}
    \mathcal{L} \supset \mathcal{G}_{R}\left(g_{w}/\sqrt{2}\right)\left[\left(U_{R}^{\mathscr{\ell}}\right)^{\dagger}U_{R}^{\nu}\right]^{ll'}\bar{N}_{R}^{l}\gamma^{\mu}l_{R}^{l'}W_{\mu}^{-} + \mathrm{h.c.,}
    \label{eq:lagrangian}
\end{equation}
and the corresponding Feynman diagram is shown in Fig. \ref{fig:feynman}. In this equation, $l_{R}$ and $N_{R}$ are right-handed mass eigenstates, $W_\mu^{-}$ is the $W$ boson field, $(U_{R}^{\mathscr{\ell}})^{\dagger}U_{R}^{\nu}$ is a $3 \times 3$ flavor mixing matrix for right-handed states, $\mathcal{G}_{R}$ is the effective coupling vertex constant, and $g_{w}$ relates to the Fermi constant through $G_{F}/\sqrt{2} = g_{w}^{2}/(8M_{W}^{2})$.

\begin{figure}
    \centering
    \includegraphics[width=0.8\columnwidth]{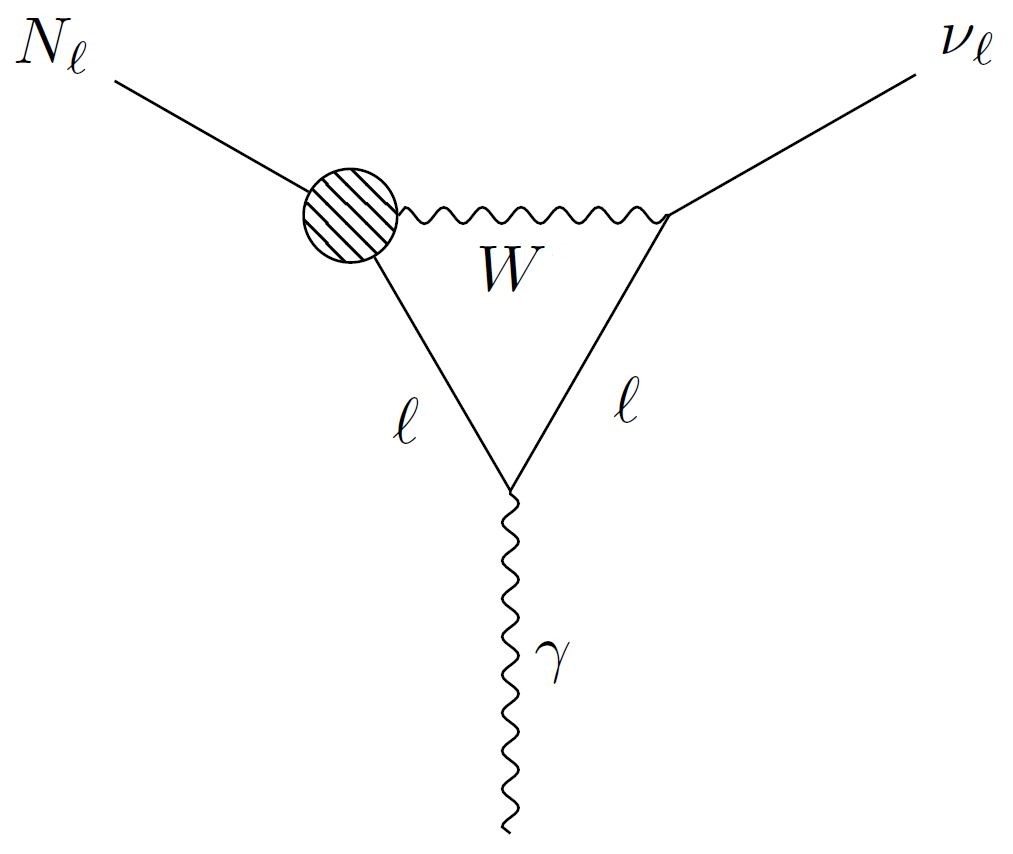}
    \caption{Effective interaction vertex between an RHDMF, a left-handed SM neutrino, and a photon. Such a process occurs in the electromagnetic channel of the coupling, and the loop involves a $W^{+}$ boson and two electrons to conserve the charge and leptonic number.}
    \label{fig:1_loop_feynman}
\end{figure}

The right-handed dark matter fermions in Eq. (\ref{eq:lagrangian}) can decay radiatively into a left-handed standard model neutrino and a photon, an electromagnetic process depicted in Fig. \ref{fig:1_loop_feynman}. The effective operator describing this decay is:

\begin{equation}
    \hat{\mathcal{O}} = \left(U^{\nu}_{L}U^{\mathscr{\ell}}_{L}\right)^{ll'}\bar{\nu}_{L}^{l}\Lambda^{\mu}_{l'}N_{R}^{l'}A_{\mu} + \mathrm{h.c.}
    \label{eq:effective_current_operator}
\end{equation}
Here, $A_\mu$ is the electromagnetic four-vector and $U^{\nu}_{L}U^{\mathscr{\ell}}_{L}$ is the Pontecorvo-Maki-Nakagawa-Sakata (PMNS) mixing matrix, which suppresses flavor oscillations in the right-handed sector. In other words, right-handed neutrinos will be considered as belonging to a diagonal representation of pure flavor states. Also, this matrix associates to the SM vertex $\bar{\nu}^{l}\gamma^{\mu}P_{L}\mathscr{\ell}^{l}W_{\mu}^{-}$ in the loop. The quantity $\bar{\nu}_{L}^{l}\Lambda^{\mu}_{l'}N_{R}^{l'}$ represents an effective current with a momentum space vertex:

\begin{equation}
    \Lambda^{\mu}_{l'}(\mathfrak{q}^{2}) = i\frac{eg_{w}^{2}\mathcal{G}_{R}m_{l'}}{16\pi^{2}}\left[\left(C_{0} + 2C_{1}\right)p^{\mu}_{\chi} + \left(C_{0} + 2C_{2}\right)p'^{\mu}\right],
\end{equation}
and the effective operator given in Eq. (\ref{eq:effective_current_operator}) is a representation of an electromagnetic property of an SM neutrino and an RHFDM coupling to a photon, which is derived from the effective right-handed current coupling in Eq. (\ref{eq:lagrangian}). The kinematics quantities $p_\chi^\mu$ and $p'^\mu$ are the incoming RHDMF and outgoing neutrino four-momenta, respectively. The squared four-momentum transfer is $\mathfrak{q}^{2} = (p_\chi^\mu - p'^\mu)^{2} = (k'^\mu - k^\mu)^{2}$, which differs from the spatial momentum transfer squared $q^2$ defined earlier. The coefficients $C_i$ ($i=0,1,2$) are the three-point Passarino-Veltman functions \cite{Passarino1979}, depending on $\mathfrak{q}^{2}$ as a variable. For these intrincated functions we used the tool \texttt{Package-X} \cite{Patel2017} to perform the corresponding calculations. 

Although the nature of the RHDMF allows it to be of any flavor, in this work we concentrate on electron-type flavor, and leave the analysis for the other types for a future work. Hence, it is possible to constrain the coupling constant of the interaction, $\mathcal{G}_{R}$, for $l = \mathrm{e}$ using the age of the Universe, which is $\sim 4.4\cdot 10^{17}$ s. Supposing the lifetime of the particle is greater than this age, this computation gives the constrain:

\begin{equation}
    \mathcal{G}_{R} \lesssim 3.1\cdot 10^{-6}\left(\frac{200\ \mathrm{keV}}{m_{\chi}}\right)^{3/2} \label{eq:G_R_e}.
\end{equation}

In the seesaw model involving neutral leptons, the radiative decay we are considering here is a subdominant process, while the dominant one is the decay $N_{l} \rightarrow \nu_{l} + \bar{\nu}_{\alpha} + \nu_{\alpha}$ \cite{Barger1995, Abazajian2001}. Since in the context of the model used in the present work, this decay channel is not present, the radiative decay $N_{l} \rightarrow \nu_{l} + \gamma$ is the dominant one. Furthermore, there is another decay channel involving electrons, which is $N_{l} \rightarrow e^{-} + e^{+} + \nu_{l}$ that is allowed to occur. But as shown in \cite{Barger1995}, the radiative decay can dominate over it.

\subsubsection{Interaction Lagrangian}

\begin{figure}
    \centering
    \includegraphics[width=0.7\columnwidth]{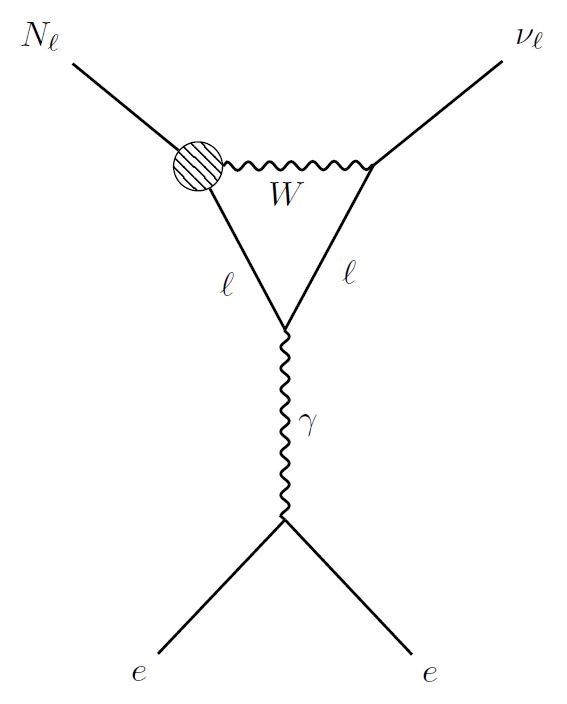}
    \caption{Inelastic scattering in the electromagnetic channel between the DM with an electron, generating two SM particles as an output, an electron neutrino, and an electron. In this diagram, the time flows to the right.}
    \label{fig:em_channel}
\end{figure}

Considering the radiative decay channel for RHDMF decay, the effective interaction Lagrangian between an electron and dark matter for an interaction as the one shown in Fig. \ref{fig:em_channel} is \cite{Shakeri2020}:

\begin{equation}
    \mathcal{L}_{\mathrm{eff}} \propto \Big[\bar{e}_{L}\gamma_{\mu}e_{L}\Big]\Big[\bar{\nu}^{e}_{L}\Lambda^{\mu}_{e}N_{R}\Big],
\end{equation}
where $N$, $\nu^{e}$, and $e$ are spinor fields for the DM, the SM electron neutrino, and electron respectively. Also, $\Lambda^{\mu}_{e}$ stands for the vertex of the effective current represented by the diagram in Fig. \ref{fig:1_loop_feynman}, for the case of the electron family, the one considered in this work. This electromagnetic channel involving a one-loop scattering process has a squared matrix element \citep[derived in][]{Shakeri2020} with the form:


\begin{equation}
    \big|M(E_{R}, \mathfrak{q}^{2}, v, \mathrm{cos}\ \theta)\big|^{2} = A\sum_{i = 0}^{10} B_{i}(E_{R}, \mathfrak{q}^{2}, \mathrm{cos}\ \theta)v^{i}.
    \label{eq:M^2}
\end{equation}
The quantity in Eq. (\ref{eq:M^2}) splits into polynomial terms in the DM velocity $v$. With the exception of $B_{0}$ and $B_{10}$, each $B_{i}$ function depends on the recoil energy $E_{R}$, squared four-momentum transfer $\mathfrak{q}^{2}$, and the scattering angle $\theta$ (angle between incoming DM and outgoing neutrino momenta). The constant $A = 8[(e^{2}g_{w}^{2}\mathcal{G}_{R})/(16\pi^{2})]^{2} m_{l}^{2}$ with $m_{l} = m_{e}$ being the virtual lepton mass in the loop, since it is an SM lepton and we are considering just the case of $l = \mathrm{e}$. The squared matrix element plays a role in the event rates formula derived in the next section.

\begin{figure*}
    \centering
        \includegraphics[width=0.495\linewidth]{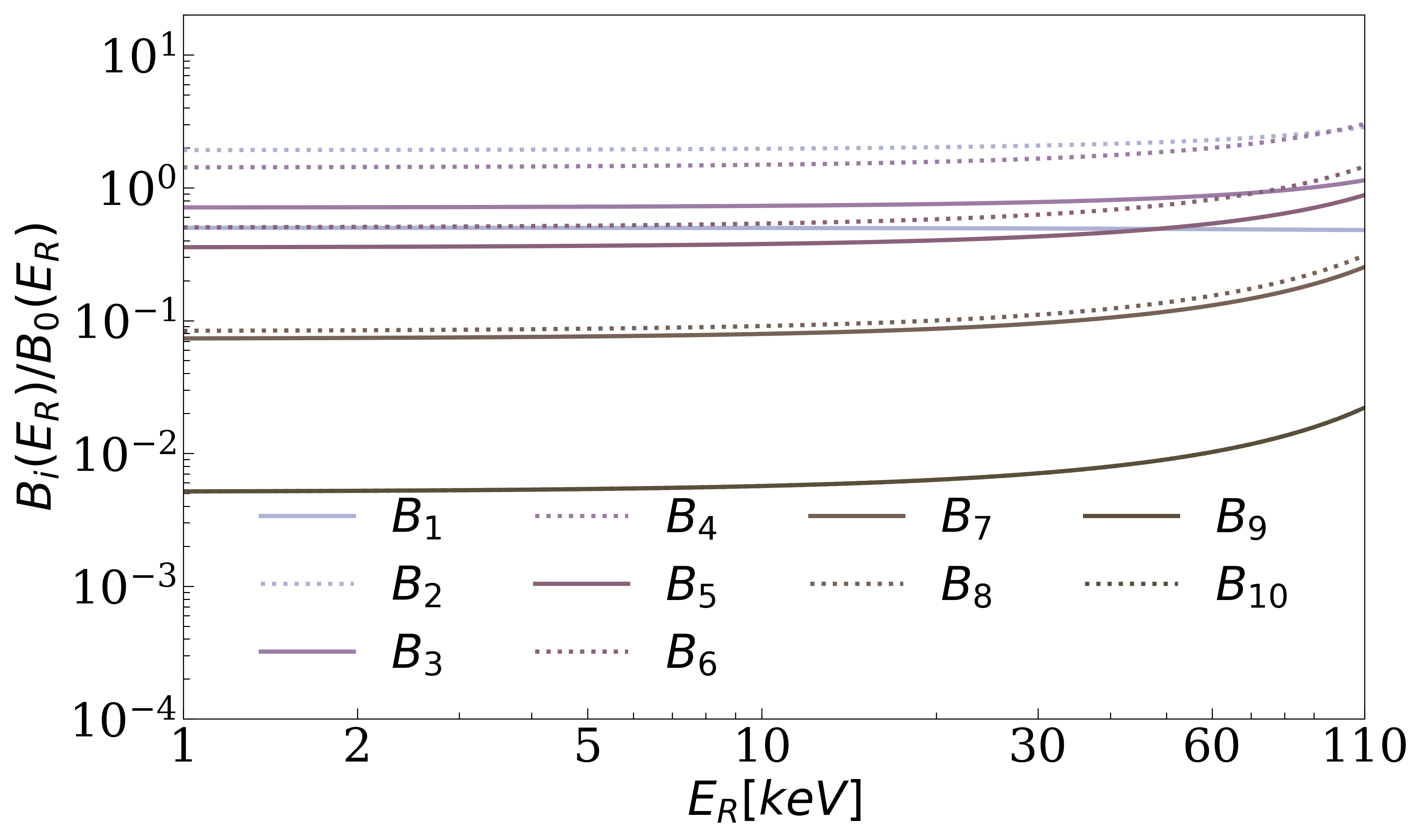}
    \hfill
        \includegraphics[width=0.495\linewidth]{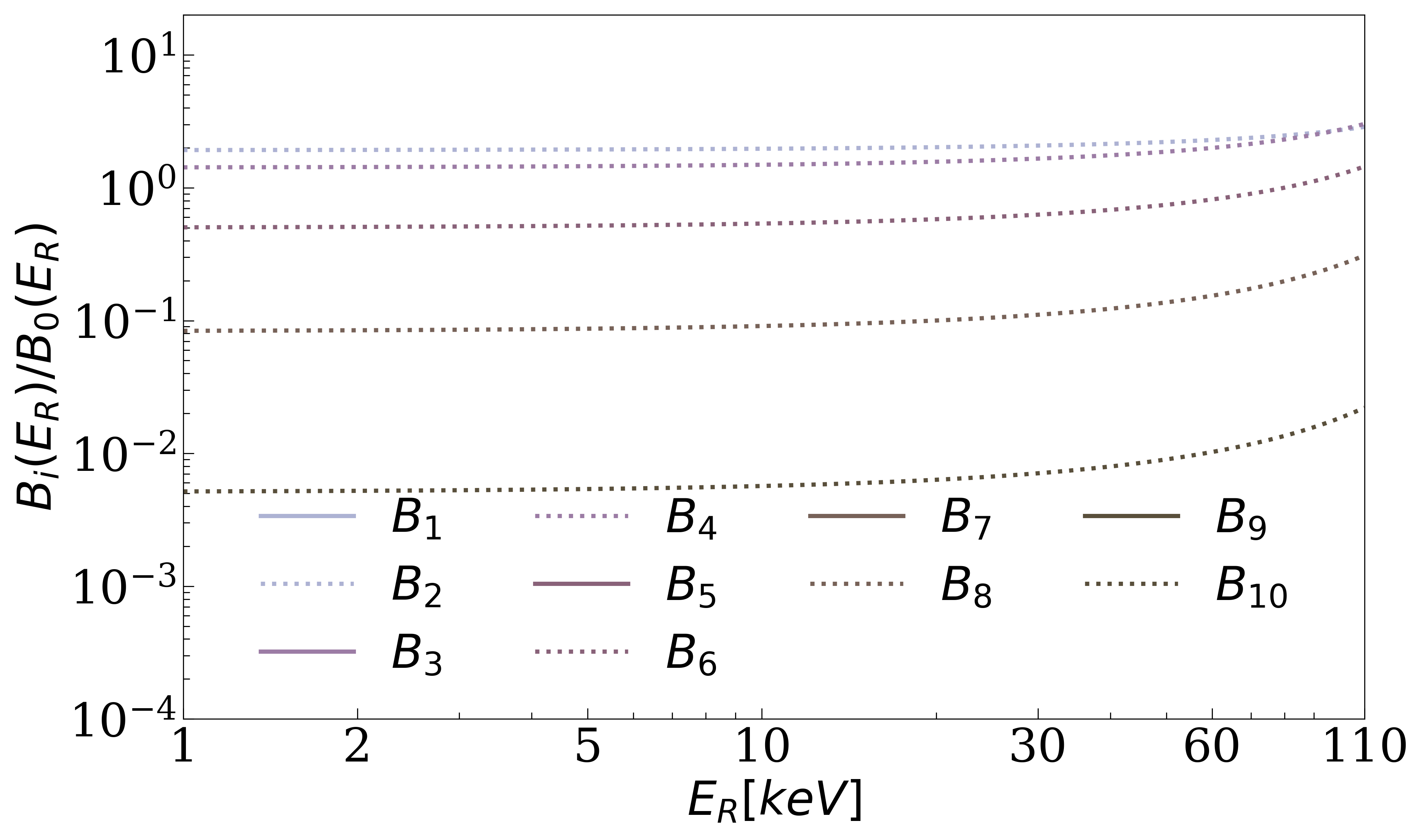}
    \caption{$B$ coefficients of the decomposition of Eq (\ref{eq:M^2}) computed for the $4d$ shell and for a DM particle mass of $m_{\chi} = 200$ keV. Left plot: functions for $\mathrm{cos}\ \theta = -\frac{1}{2}$. Right plot: functions for $\mathrm{cos}\ \theta = \frac{1}{2}$. Note that in the right plot the functions with odd index are not displayed. This is because they are negative for $\mathrm{cos}\ \theta = \frac{1}{2}$, but the ratio $B_{i}/B_{0}$ is still negligible in this case.}
    \label{fig:b_functions}
\end{figure*}

The analysis of $B_{i}$ functions shows (Fig. \ref{fig:b_functions}) they have comparable magnitudes when contrasted to $B_{0}$. Combined with powers of typical DM speeds $v \sim 10^{-3}$ in natural units, only the $B_{0}$ term dominates. This yields:

\begin{equation}
    \big|M(E_{R}, \mathfrak{q}^{2})\big|^{2} = A B_{0}(E_{R}, \mathfrak{q}^{2}).
\end{equation}
The function depending on the recoil energy and the squared four-momentum transfer, has the form:

\begin{widetext}
    \begin{equation}
        \begin{split}
            B_{0}(E_{R}, \mathfrak{q}^{2}) = &\ \frac{m_{\chi}}{\mathfrak{q}^{4}}(E_{B}^{nl} - E_{R} + m_{\chi})\Big[2(C_{0} + 2C_{2})^{2}(E_{B}^{nl} - E_{R} + m_{\chi})^{2}(E_{B}^{nl} + m_{l})(E_{B}^{nl} + m_{\chi} + m_{l}) + \\ 
            & + (C_{0} + 2C_{1})^{2}m_{\chi}^{2}\big(E_{B}^{nl}(E_{R} + m_{l}) + m_{l}(E_{R} + 2m_{l})\big) + 2(C_{0} + 2C_{1})(C_{0} + 2C_{2})\cdot \\
            & \cdot m_{\chi}(E_{B}^{nl} - E_{R} + m_{\chi})\big((E_{B}^{nl})^{2} + (E_{B}^{nl} + m_{l})(m_{\chi} + 2m_{l})\big)\Big].
        \end{split}
    \end{equation}
\end{widetext}

\begin{figure}[tbp]
    \centering
    \includegraphics[width=\columnwidth]{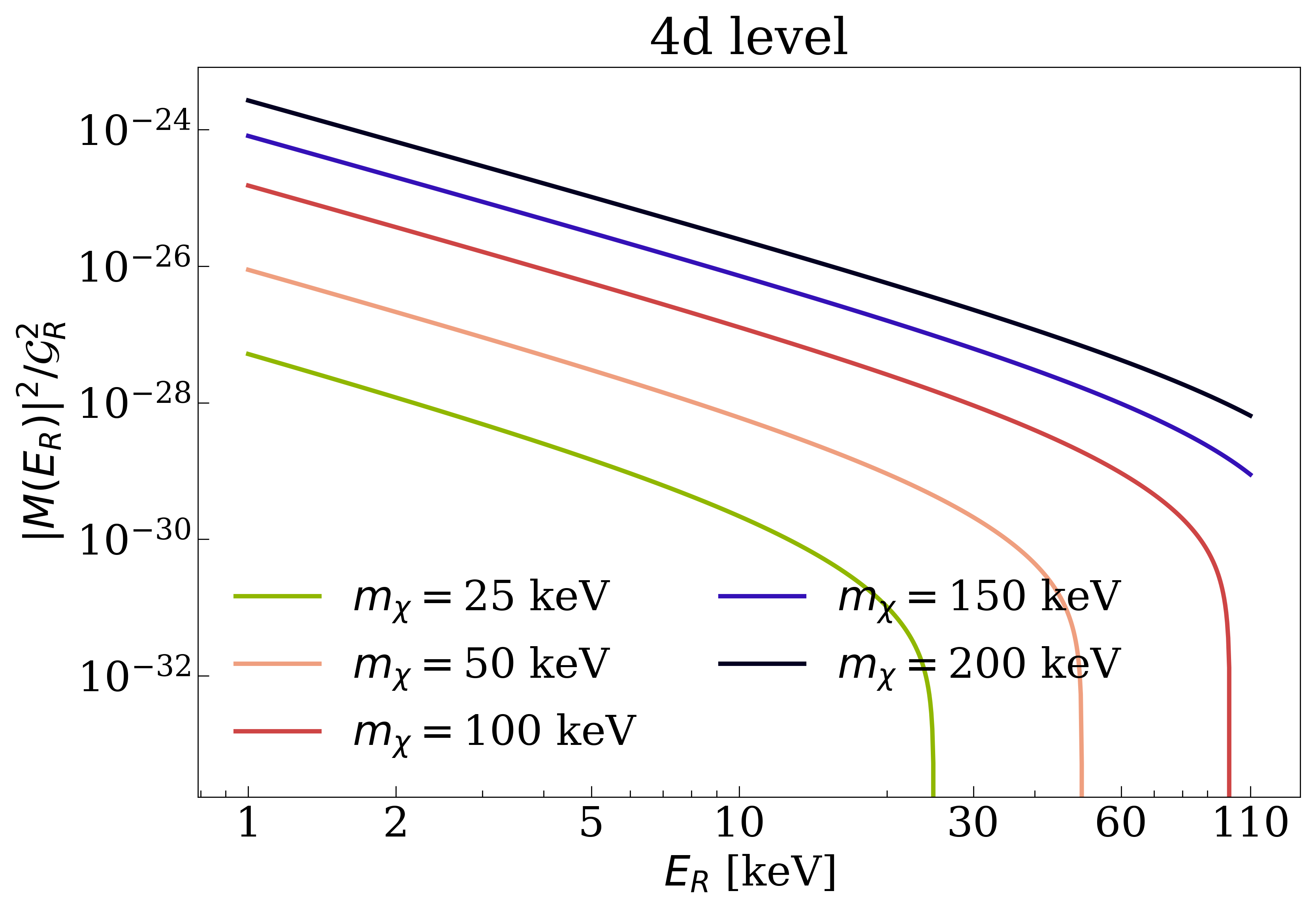}
    \caption{Squared matrix element of the four-fermion interaction between the incoming RHDMF and the bound electron. It is displayed as a function of the recoil energy for five different DM particle masses, $m_{\chi} = 200,\ 150,\ 100,\ 50,\ 25$ keV, each case corresponding to the $4d$ shell of the xenon atom.}
    \label{fig:m_squared}
\end{figure}

Since $\mathfrak{q}^{2}$ exhibits a quadratic relationship with $E_{R}$, derived in App. \ref{ap:squared_four_momentum_transfer} with $\mathfrak{q}^{2} = (E_{R} - E_{B}^{nl})^{2} - 2m_{e}E_{R}$, the squared matrix element simplifies to depend only on recoil energy:

\begin{equation}
    \big|M(E_{R})\big|^{2} = A B_{0}(E_{R}).
    \label{eq:m^2_final}
\end{equation}
A representation of this function for three different DM masses and for the $4d$ shell of the xenon atom is shown in Fig. \ref{fig:m_squared}.

It is noteworthy that the squared matrix element, though derived for free scattering $\chi + e^{-} \rightarrow \nu + e^{-}$, incorporates the bound electron through the four-momentum transfer $\mathfrak{q}^{2} = (E_{R} - E_{B}^{nl})^{2} - 2m_{e}E_{R}$, which depends on the atomic binding energy. Thus $|M|^{2}$ depends on the atomic potential indirectly, while the ionization form factor below captures atomic structure more explicitly. Both contributions together determine the event rate morphology under the approximation of Eq. (\ref{eq:energy_conservation_approximation}).

\subsection{Electron recoil process}

\begin{figure}
    \centering
    \includegraphics[width=\columnwidth]{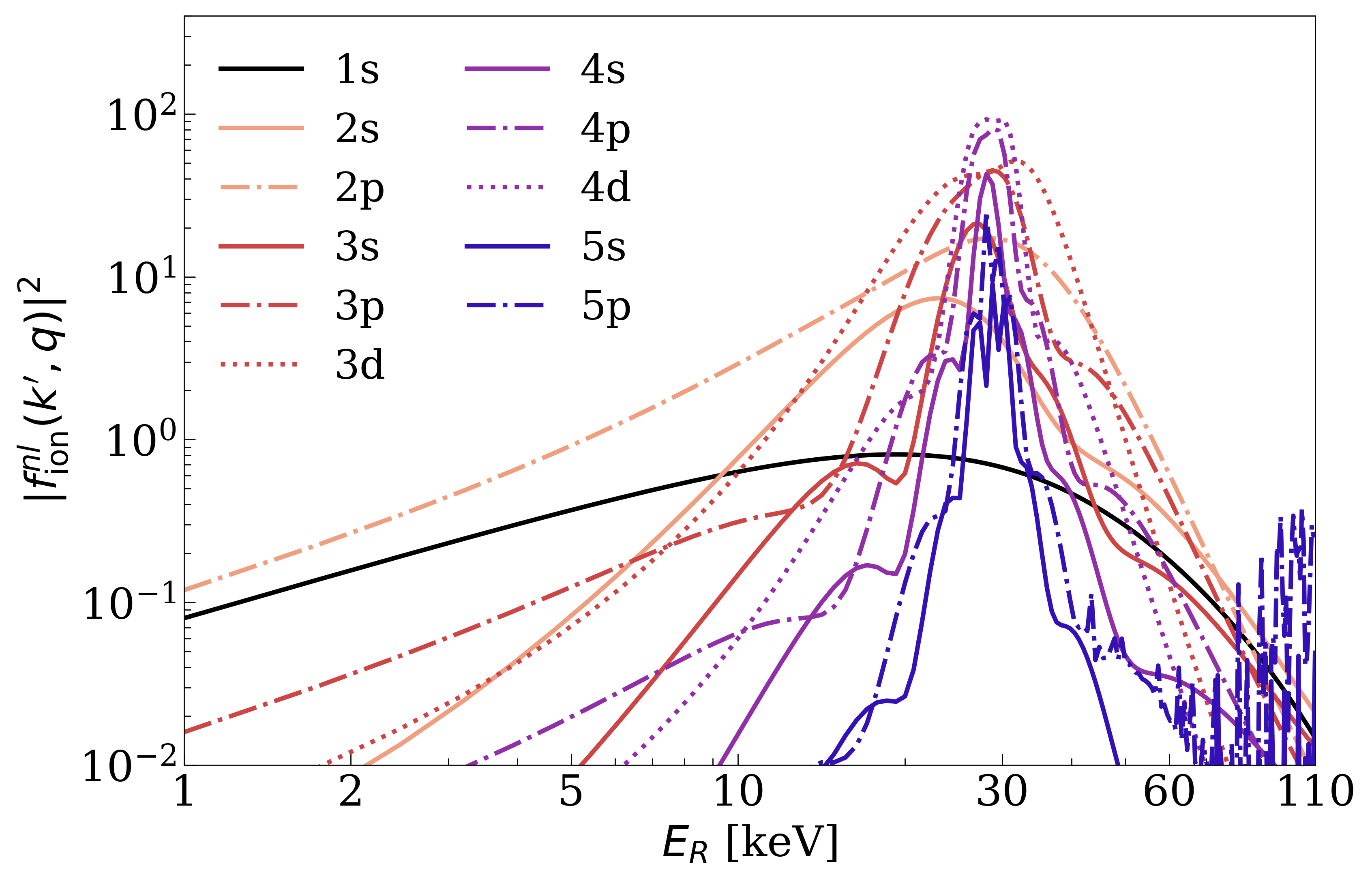}
    \caption{Ionization form factors computed for all the energy levels of the xenon atom as a function of the recoil energy. This is for $m_{\chi} = 200$ keV, which plays a role in the energy conservation equation. Different colors indicate different $n$ quantum numbers, while different line styles separate the quantum number $l$.}
    \label{fig:f_ion_all}
\end{figure}

We need to quantify the rate of interactions the RHDMF particles have with the material of the detector. This is possibly done with the definition of event rates. This quantity indicates the rate of events on the detector per unit of time, mass of the detector material, and per unit of energy. Based on the framework derived on App. \ref{ap:event_rates}, the differential event rate for RHDMF scattering with an electron bound in xenon shell $(n, l)$ is:

\begin{equation}
    \frac{dR_{\mathrm{ion}}^{nl}}{dE_{R}} = \frac{A\rho_{\chi}}{64\pi m_{\chi}^{2}m_{e}^{2}}B_{0}(E_{R})\mu_{0}\frac{q}{E_{R}}\Big|f_{\mathrm{ion}}^{nl}(k', q)\Big|^{2}.
    \label{eq:events_rate_nl}
\end{equation}
Here, $\rho_{\chi} = 0.53$ GeV cm$^{-3}$ (or $4.11 \times 10^{-18}$ keV$^{-4}$ in natural units) is the local dark matter density from \cite{Arguelles2023a}. On the other hand, the value of $q$ is related to the recoil energy based on the energy conservation condition given in Eq. (\ref{eq:energy_conservation_approximation}). The event rate depends on two key physical quantities. First, the ionization form factor for an electron in shell $(n, l)$ with post-ionization momentum $k' = \sqrt{2m_{e}E_{R}}$ (see App. \ref{ap:event_rates} for further details):

\begin{widetext}
    \begin{equation}
        \Big|f_{\mathrm{ion}}^{nl}(k', q)\Big|^{2} = \frac{4k'^{3}}{(2\pi)^{3}}\sum_{l'=0}^{\infty}\sum_{L=|l-l'|}^{l+l'}(2l + 1)(2l' + 1)(2L + 1) \begin{pmatrix}
            l & l' & L \\
            0 & 0 & 0
        \end{pmatrix}
        \left|\int dr r^{2}R_{k'l'}(r)R_{nl}(r)j_{L}(qr)\right|^{2}.
        \label{eq:f_ion}
    \end{equation}
\end{widetext}
where the round brackets denote the Wigner $3j$ symbol, $R_{k'l'}(r)$ is the wave function of the outgoing electron, and $R_{nl}(r)$ is the bound electron wave function. The first wave function is based on the assumption of a radial Schrödinger equation with a central potential $Z_{\mathrm{eff}}/r$. The bound electron is described by Roothaan-Hartree-Fock ground state wave functions as linear combinations of Slater-type orbitals. Explicit expressions and plots of radial wave functions appear in App. \ref{ap:radial_wave_functions}. Figure \ref{fig:f_ion_all} shows ionization form factors as a function of the recoil energy for all xenon atomic shells.

The second key quantity appearing indirectly on Eq. (\ref{eq:events_rate_nl}) is the squared matrix element $\big|M(E_{R}, v, \mathrm{cos}\ \theta)\big|^{2}$, which encapsulates the quantum field theory of the RHDMF-electron interaction. Integration over the DM speed distribution in App. \ref{ap:event_rates} introduces the zeroth moment $\mu_{0}$, which for a normalized distribution, it is $\mu_{0} = 1$, preserving the recoil energy-dependence through the $B_{0}$ function. Another comment worth mentioning is that the physics of the interaction and the atomic physics are almost independent. The squared matrix element depends on the potential of the xenon atom through the value of $\mathfrak{q}^{2} = \left(E_{R} - E_{B}^{nl}\right)^{2} - 2m_{e}E_{R}$, while the ionization form factor is not a function of the Feynman amplitudes. Yet, both physical magnitudes conform the essence of the event rates morphology under the assumption suggested in \cite{Dror2021} that allows to write a simpler energy conservation equation such as that of Eq. (\ref{eq:energy_conservation_approximation}).

Modeling the events-rate involves the theoretical recoil energy $E_{R}$, while the experiments report their results using another recoil variable. This is related with the calibration of the detectors, and we call it `visible' recoil energy $E$. Considering that the corresponding XENONnT experiment has a relative efficiency $f = E/E_{R}$ \cite{Lewin1996}, the theoretical event rates spectra has to be replaced with the experimental one, which is:

\begin{equation}
    \frac{dR_{\mathrm{ion}}^{nl}}{dE} = \frac{1}{f}\left(1 - \frac{E}{f}\frac{df}{dE}\right)\frac{dR_{\mathrm{ion}}^{nl}}{dE_{R}}.\label{eq:events_rate_nl_visible_energy}
\end{equation}
Then, to compute the total event rates the quantity defined on Eq. (\ref{eq:events_rate_nl_visible_energy}) has to be summed up for all the energy levels:

\begin{equation}
    \frac{dR_{\mathrm{ion}}}{dE} = C N_{T}\sum_{nl}\frac{dR_{\mathrm{ion}}^{nl}}{dE}.
    \label{eq:total_events_rate}
\end{equation}
Here, $N_{T} = 4 \times 10^{27}$ tonne$^{-1}$ is the number of xenon atoms per unit detector mass, and $C = 4.79 \times 10^{25}$ year$^{-1}$ keV$^{-1}$ is a unit conversion factor to match literature conventions in natural units.

\subsection{Total cross section}

With the squared matrix element determined, we can compute the differential cross section for $\chi + e^{-} \rightarrow \nu + e^{-}$ with the initial electron bound to a xenon atom. In the laboratory frame where the initial electron is at rest ($\vec{k} = 0$), so $E_k = m_e + E_B^{nl}$, and using the non-relativistic DM limit ($E_\chi \approx m_\chi$) with $p_\chi \ll p'$, the differential cross section reads:

\begin{equation}
    \frac{d\sigma^{nl}_{\mathrm{LAB}}}{d\Omega} = \frac{\big(64\pi^{2}m_{\chi}m_{e}v\big)^{-1}\big|M(E_{R})\big|^{2}}{1 + \sqrt{1 + \frac{4\big(m_{\chi} + m_{e} + E_{B}^{nl}\big)^{2}m_{e}^{2}}{\big(m_{\chi} + E_{B}^{nl}\big)^{2}\big(m_{\chi} + 2m_{e} + E_{B}^{nl}\big)^{2}}}}.
\end{equation}
Since $|M|^{2}$ is independent of the scattering angle, as stated in Eq. (\ref{eq:m^2_final}), integration over solid angle is straightforward, yielding:

\begin{equation}
    \sigma^{nl}_{\mathrm{LAB}}(E_{R}) = \frac{\big(16\pi m_{\chi}m_{e}v\big)^{-1}AB_{0}(E_{R})}{1 + \sqrt{1 + \frac{4\big(m_{\chi} + m_{e} + E_{B}^{nl}\big)^{2}m_{e}^{2}}{\big(m_{\chi} + E_{B}^{nl}\big)^{2}\big(m_{\chi} + 2m_{e} + E_{B}^{nl}\big)^{2}}}}.
\end{equation}
This represents the cross section for inelastic RHDMF scattering with an electron in xenon shell $(n,l)$ in the non-relativistic limit under the approximation of Eq. (\ref{eq:energy_conservation_approximation}). The total cross section summing all shells is:

\begin{equation}
    \sigma_{\mathrm{LAB}}(E_{R}) = \sum_{nl} \sigma^{nl}_{\mathrm{LAB}}(E_{R}).
\end{equation}

It is worth noting that, due to the zeroth-order approximation of the DM energy $E_{\chi}$, the total cross section depends only on the recoil energy of the outgoing electron and on the speed of the incoming DM particle. The remaining quantities are fixed by theory parameters.

\section{Results}\label{sec:results}

We present predictions for fermionic dark matter absorption by electrons bound in xenon atoms, including ionization event rates and exclusion regions for coupling parameters.

\subsection{Event rates}

\begin{figure*}
    \centering
        \includegraphics[width=0.495\linewidth]{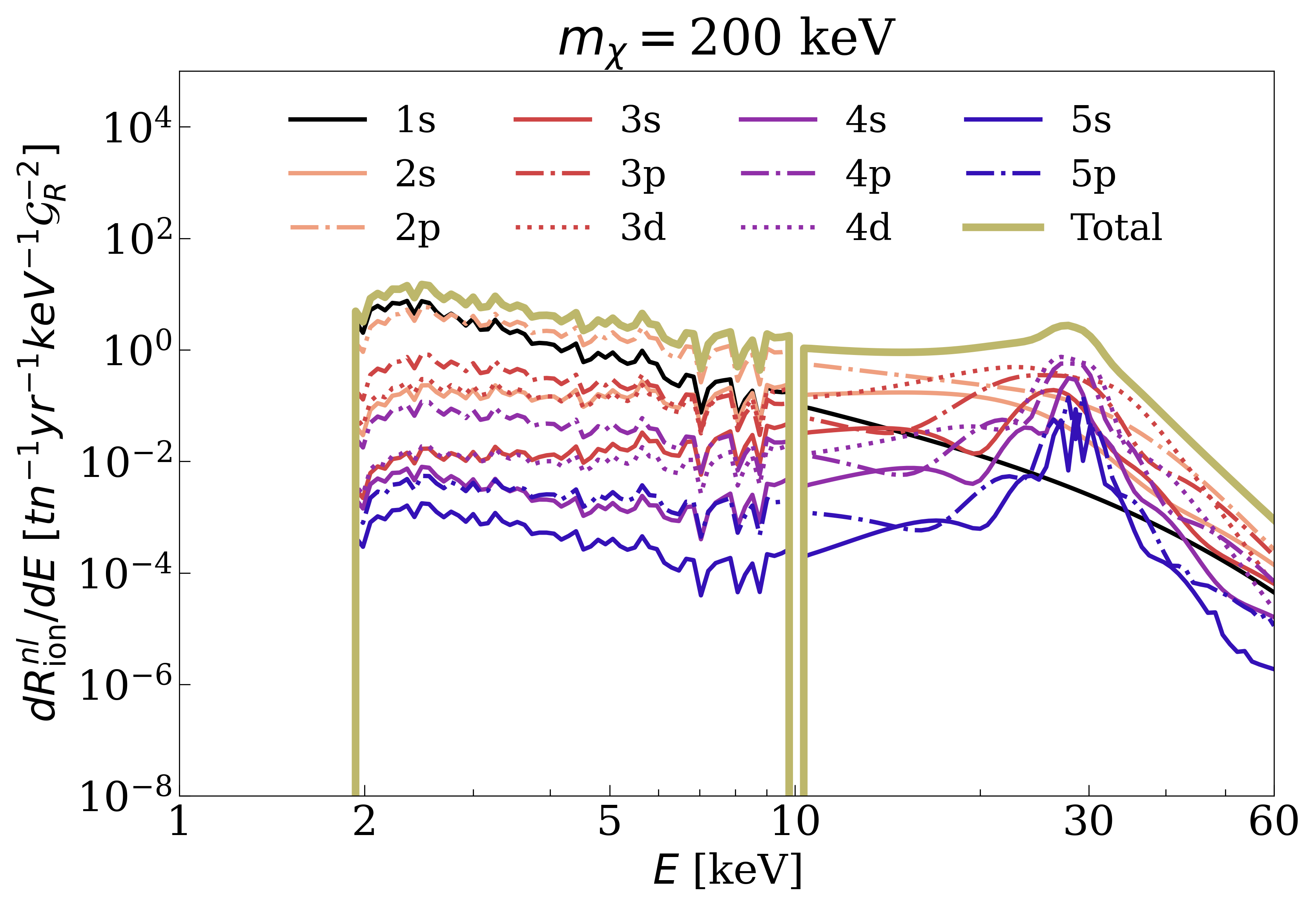}
    \hfill
        \includegraphics[width=0.495\linewidth]{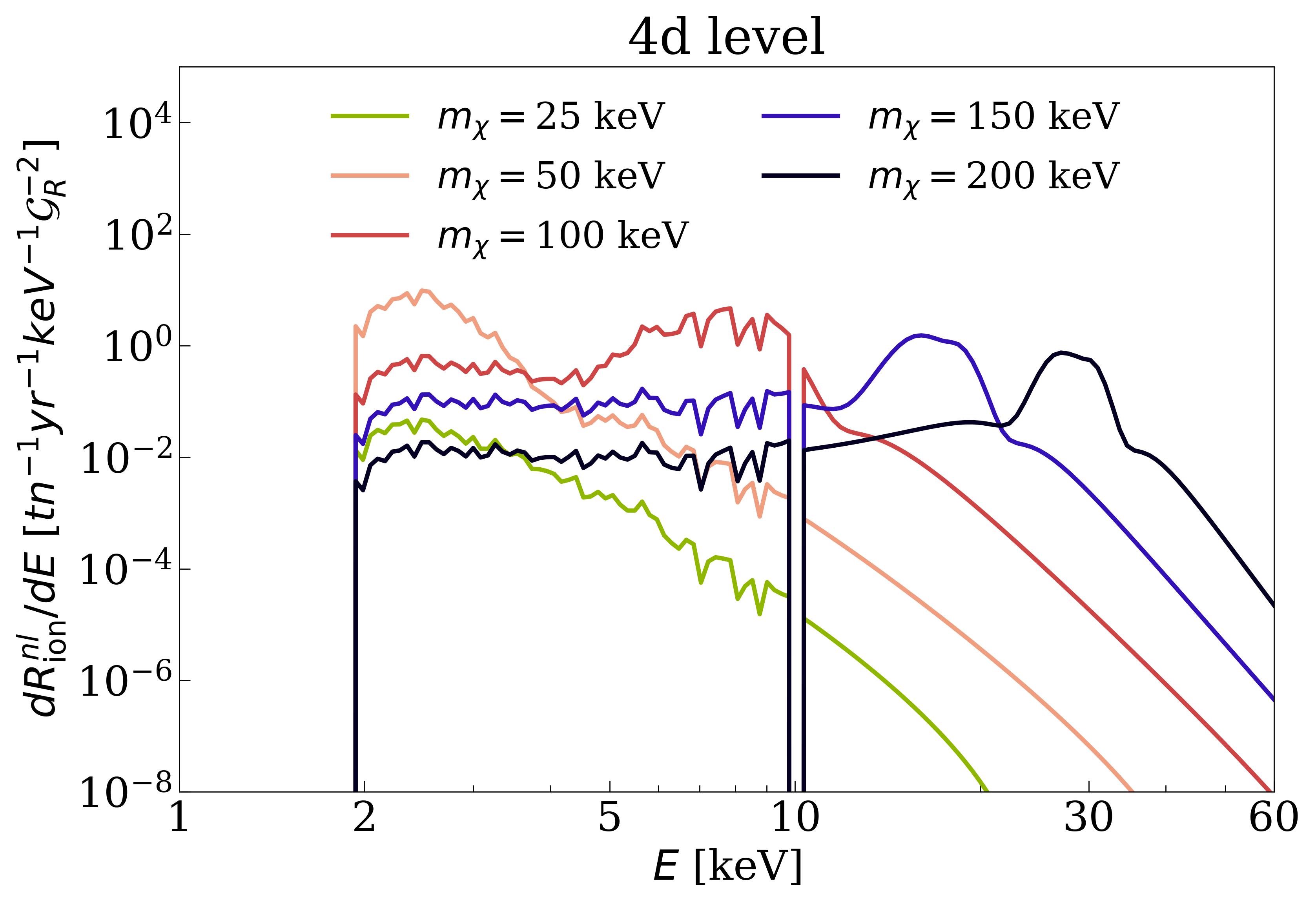}
    \caption{Event rates for different shells and masses. Left: event rates for every shell level of the xenon atom, for fixed DM mass of $m_{\chi} = 200$ keV. In khaki, the total contribution, computed using Eq. (\ref{eq:total_events_rate}). Right: event rates corresponding to the $4d$ quantum level, but for different DM masses, as indicated in the legend.}
    \label{fig:events_rate}
\end{figure*}

Applying Eq. (\ref{eq:events_rate_nl_visible_energy}) for each xenon shell yields the predicted interaction rate. Figure \ref{fig:events_rate} displays the behavior for different mass values and atomic levels. The left panel shows individual shell contributions for fixed DM mass $m_\chi = 200$ keV, with the total summed over all shells in khaki. The rate exhibits rapid variation at low energies, smoothing at energies above $\sim 10$ keV with a peak near $E \sim 30$ keV from the ionization form factor (Fig. \ref{fig:f_ion_all}). Above this peak, the rate decreases several orders of magnitude.

The local minimum near $E \sim 10$ keV reflects a discontinuity in the XENONnT efficiency \citep[Fig. 1 in][]{XenonCollaboration2022}, corresponding to the still-blinded WIMP search region. The rate increase at lower visible energies reflects the squared matrix element behavior (Fig. \ref{fig:m_squared}). The right panel shows rates for the $4d$ shell with varying DM mass. Peaks at $\sim 3$ keV, $\sim 8$ keV, $\sim 15$ keV, and $\sim 30$ keV mark ionization form factor enhancements. In addition, a well established behaviour is the increase in the event rates spectra for lower DM particles masses and towards smaller visible energies, also reflecting the form factor's mass-dependence at small recoils. The maximum of each peak is even an increasing function for decreasing values of the incoming RHDMF mass.

Another interesting and expected feature of the event rates spectra when analysing the left plot of Fig. \ref{fig:events_rate} is the fact that, for small visible recoil energies, the levels dominating the contribution to the total, are the innermost ones. This stems from, when an incoming DM particle collides with an inner electron and ionizes the atom, this free electron has to go through the different external shells. In its way, the outgoing charged particle will ionize other electrons, repeating the same process for the new ionized electron. This cascade effect repeats until the outermost electrons surpass the most external shell, reaching the detector with a small kinetic energy. Since this whole process was triggered by an ionization taking place in an internal energy level, this kind of interactions enhance the event rates spectra for small visible recoil energies and most internal xenon shells.

\subsection{Exclusion regions}

\begin{figure*}
    \centering
        \includegraphics[width=0.495\linewidth]{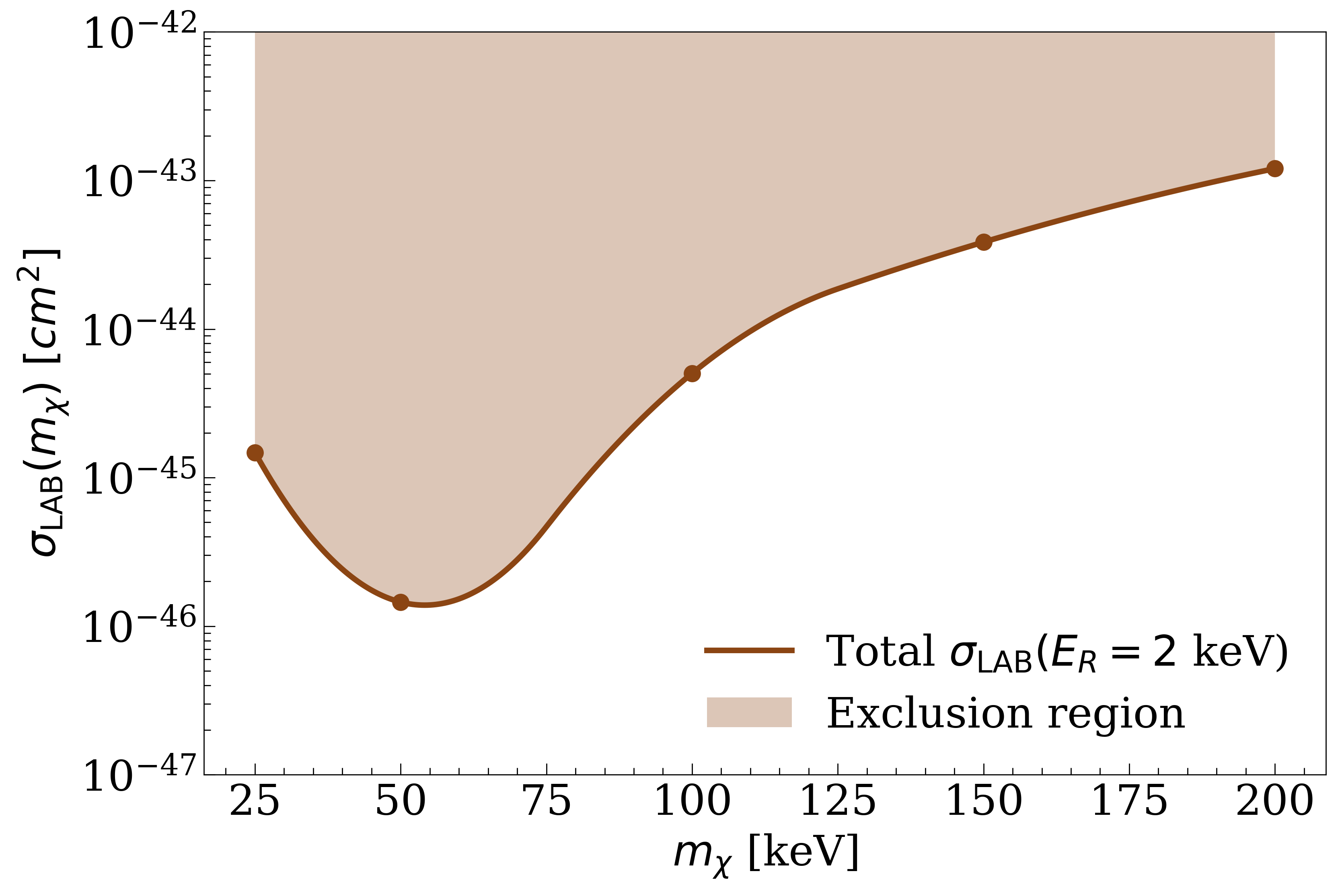}
    \hfill
        \includegraphics[width=0.495\linewidth]{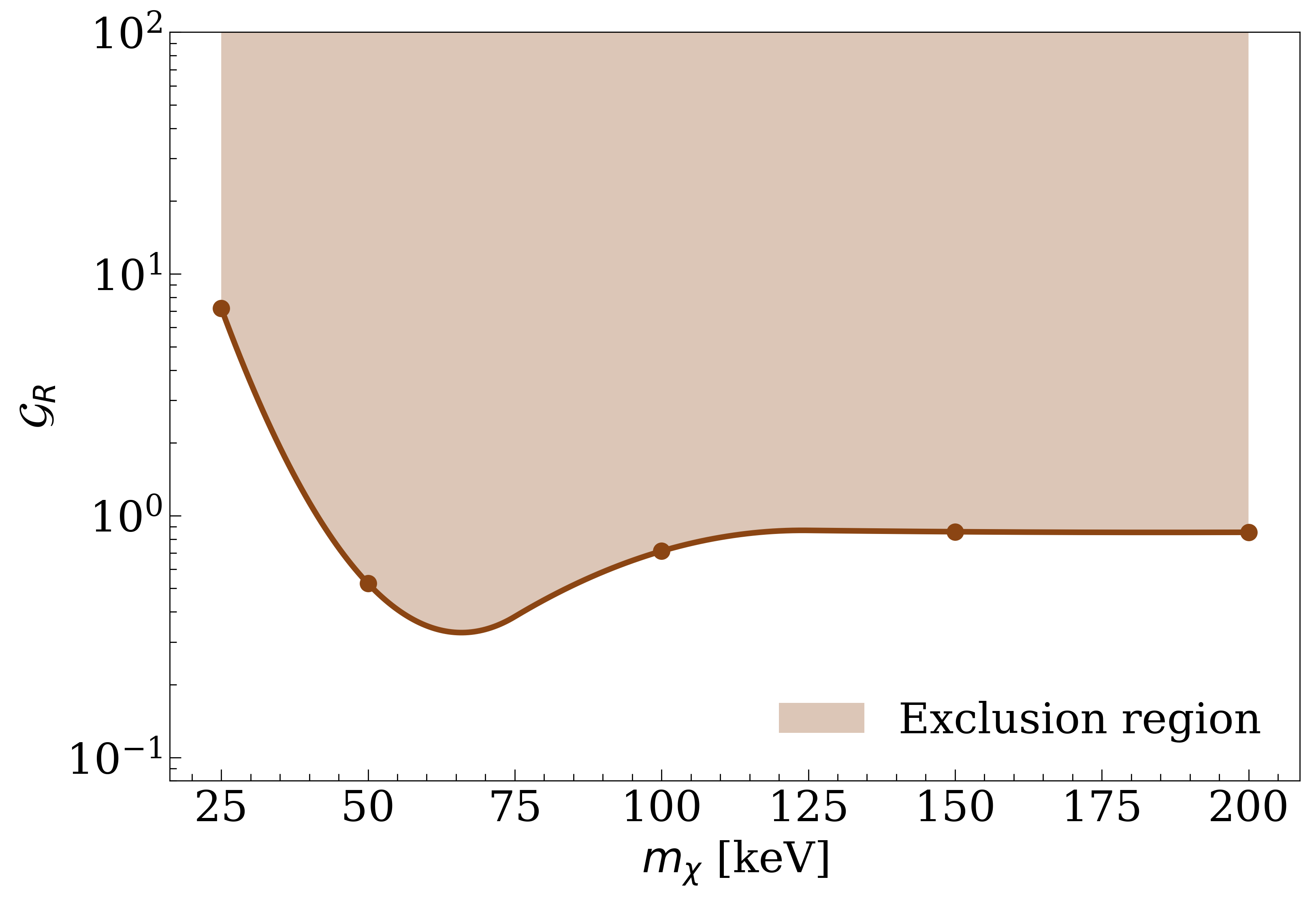}
    \caption{Exclusion regions for the interaction model between an RHDMF and electrons bound in xenon atoms. Left: excluded region in the plane of $\sigma_{\mathrm{LAB}}$ (the total cross section) versus $m_{\chi}$. Right: excluded region in the plane of $\mathcal{G}_{R}$ (the coupling constant) versus $m_{\chi}$. In both plots the points correspond either to the total cross section or the coupling constant for the mass values under consideration, while those points are quadraticaly interpolated to smooth the limiting curve.}
    \label{fig:exclusion_regions}
\end{figure*}

Using the total predicted event rates for XENONnT is possible to set exclusion regions for the interaction model between an RHDMF and electrons bound in xenon atoms. This is done by constraining the parameter space based on the comparison between our predictions and the experimental data. The last one is the observed spectra of events in XENONnT, which rules out any positive DM detection since the fit spectra can be completely explained through background processes. By making the predicted total event rates spectra to be under the determined upper limit in \cite{XenonCollaboration2022}, we can exclude regions of the parameter space. The exclusion regions are shown in Fig. \ref{fig:exclusion_regions} for two different representations. The left panel shows the excluded region in the plane of $\sigma_{\mathrm{LAB}}$ (the total cross section) versus $m_{\chi}$, while the right panel shows it in the plane of $\mathcal{G}_{R}$ (the coupling constant) versus $m_{\chi}$. In both cases, the allowed region is below the curve, while the excluded one is above it, denoted in shaded terracotta. Despite the upper limit imposed on the coupling constant by cosmology is smaller than the one obtained by constraining the model with observed data, it has to be remarked that in both (independent) cases they are upper limits. In the case experimental accuracy of direct detection experiments keep improving, the upper limit derived with them will eventually continue decreasing, making tighter the allowed region.

It is very interesting to note the behaviour of the lower limit of the excluded region, for both panels in Fig. \ref{fig:exclusion_regions}. In the left plot, the minimum is $\sim 50$ keV for the DM particle mass, conserving the same feature for the right plot, though the trend tends to flatten for masses above this value. This is due to the fact that the larger total contribution of the predicted event rates spectra is for $m_{\chi} = 50$ keV, based on our model. Since there is no excess above the background, it is expected that for masses near $50$ keV the exclusion region is wider. While it means RHDMF masses near $50$ keV are more difficult to constrain, it is worth to mention that there are quite interesting dark matter halo astrophysical results pointing towards fermions whose mass is $\mathcal{O}(10$--$100$ keV), as detailed in \cite{Arguelles2018,Arguelles2019,Arguelles2022,Arguelles2023a,Collazo2025}.

\section{Conclusions}\label{sec:conclusions}

In this work we have described the interaction of fermionic, electrically neutral, right-handed dark matter particles with Xe atoms. The mechanism for such a process requires the treatment of interactions between right-handed DM particles and bound left-handed electrons belonging to the atomic orbits of Xe. The cross section corresponding to the ionization of Xe atoms is obtained as a function of the mass of the DM particles, of the strength of the right-left currents coupling, and the recoil energy of the outgoing electron. The calculations performed under non-relativistic approximations allow for the setting of allowed and excluded regions, determined by the dependence of the ionization rate upon the mass of the incoming DM particles, the strength of the coupling with them to electrons bound to different atomic orbits of Xe and by the energy of these orbits. These ionization rates are strongly influenced by the quantum nature of the Xe atom, reflecting the fact the target electron is a non-free particle in the process. The interaction of DM and electronic states of Xe was described as a three step process where, firstly, the incoming DM particle decays in a $W^{+}$ boson and a right-handed electron, both virtual particles of the one-loop process. Secondly, after the chiral flip of the right-handed electron induced by the fermionic propagator, the left-handed electron emits a photon, whose absortion by the bounded electron will excite the last one among different quantum levels of the xenon atom. Finally, the virtual left-handed electron will combine with the virtual $W^{+}$ to produce an outgoing electron-neutrino of the SM. From the results of the calculations where we have considered the values of the mass of DM particles in the domain 25 keV$/c^{2}$ $\leq m_\chi \leq$ 200 keV$/c^{2}$, we determined the upper limit values of the right-left coupling $\mathcal{G}_{R}$ varying from $\mathcal{O}(1)$ up to $\mathcal{O}(10)$.

The aforementioned exclusion regions were computed using the events per year, tonne, and energy reported in \cite{XenonCollaboration2022}, where the $1.16$ ton-years exposure of the experiment determined no excess above the background. Our results show the range of DM particle masses around $50$ keV$/c^{2}$ is the most confident one, excluding cross sections above $\sim 10^{-46}$ cm$^{2}$. Motivated by the absence of potential positive events and by the shape of the observed spectral profile, we did not fit the predicted signal to the observed one, but we focused our work on making the predicted event rates to be under the data. Beside this, it is still interesting the agreement between our DM mass value of the tightest constrain and those utilized on other works, such as \cite{PandaXCollaboration2022, Geng2025, PandaXCollaboration2025}. Also, the agreement on the shape of the exclusion curves is noteworthy, presenting them a valley for DM masses around $50$ keV$/c^{2}$.

As a last comment, it is worth to remark on the shape of the DM predicted spectrum, as shown in Fig. \ref{fig:events_rate}. The fact that the ionization form factor affects the event rates behaviour by introducing a DM mass-dependent peak is an advantage for future DM direct detection experiments. Despite there is still no convincing evidence of positve events, as for example reported with the ionization-only (S2-only) signals in XENONnT with a total exposure of $7.83$ ton-years in \cite{XenonCollaboration2026}, this special form of the spectra could make the difference in these kind of experiments. If confirmed the presence of positive non-background events under a certain statistical confidence, the peak in the predicted event rates could drive to determine the mass of the incoming particle. Measuring this quantity involving very precise quantum atomic physics driven by the radial wave functions like the ones considered in this work is extremely important for the future prospect of direct detection experiments.

\begin{acknowledgments}
S.C. acknowledges CONICET from Argentina and FCAGLP for the finantial and institutional support that made possible this work. S.C. and C.R.A. thanks Martín Schvellinger for fruitful discussions about the effective left-right current. S.C. thanks Roberto Morales for rewarding comments on the squared matrix element of the one-loop process. S.C, C.R.A. and O.C. thanks She-Sheng Xue for valuable discussions about the interaction model. C.R.A. thanks financial support from CONICET. O.C. is a member of the CONICET and acknowledges the PIP 2081. We thank IALP support and administrative staff for their dedicated work.
\end{acknowledgments}

\appendix
\onecolumngrid

\section{Squared four-momentum transfer}\label{ap:squared_four_momentum_transfer}

Given the momentum transfer vector $\vec{q} = \vec{p}_{\chi} - \vec{p'} = \vec{k'} - \vec{k}$, in a relativist framework it can be redefined in terms of the four-momentum transfer $\mathfrak{q}^{\mu} = p_{\chi}^{\mu} - p'^{\mu} = k'^{\mu} - k^{\mu}$. Using the nomenclature and physical context defined in Sec. \ref{sec:physical_context_and_formulae} for the momenta and energies of the electrons in the process, it can be written that $k^{\mu} = (m_{e} + E_{B}^{nl};\ \vec{0})$ and $k'^{\mu} = (m_{e} + E_{R};\ \vec{k'})$. It is worth recalling that both charged particles involved in the scattering are treated in the non-relativistic limit, so the total energy of both is the sum of their rest mass and their potential or kinetic energy.

It is straightforward to see that, in this case, the four-momentum transfer is:

\begin{equation}
    \mathfrak{q}^{\mu} = \left(E_{R} - E_{B}^{nl};\ \vec{k'}\right).
\end{equation}
This way, the contraction of the four-momentum transfer with itself can be computed, which reads:

\begin{align}
    \mathfrak{q}^{2} \equiv \mathfrak{q}_{\mu}\mathfrak{q}^{\mu} &= \mathfrak{q}^{\mu}\mathfrak{q}^{\nu}\eta_{\mu \nu}, \\
    &= \left(E_{R} - E_{B}^{nl}\right)^{2} - \vec{k'}^{2}.
\end{align}
Note that the signature convention chosen to the Minkowski metric is the one called `mostly less'. Recalling that $\vec{k'}^{2} = 2m_{e}E_{R}$, the last line above yields:

\begin{equation}
    \mathfrak{q}^{2} = \left(E_{R} - E_{B}^{nl}\right)^{2} - 2m_{e}E_{R}.
\end{equation}
This way, there is a quadratic relationship between $\mathfrak{q}^{2}$ and $E_{R}$.

\section{Event rates calculation}\label{ap:event_rates}

The event rate for electron recoils in the detector is obtained from the scattering cross section for dark matter interacting with bound electrons. We start from the general expression in Eq. (A.12) of \cite{Essig2016}:

\begin{equation}
    \sigma v_{1 \rightarrow 2}(v) = \frac{1}{16\pi m_{\chi}^{2}m_{e}^{2}}\int \frac{d^{3}q}{4\pi}\delta\left(\Delta E_{1 \rightarrow 2} + \frac{q^{2}}{2m_{\chi}} - qv\cos(\theta_{qv})\right)\Big|M(E_{R}, v, \cos\theta)\Big|^{2}\Big|f_{1 \rightarrow 2}(\vec{q})\Big|^{2}.
    \label{eq:sigma_v}
\end{equation}
The squared matrix element $\big|M(E_{R}, v, \cos\theta)\big|^{2}$ depends on the recoil energy $E_{R}$, but energy conservation from Eq. (\ref{eq:energy_conservation}) allows us to express $E_{R}$ in terms of the momentum transfer $q$. The matrix element is averaged over initial spin states and summed over final states, while the delta distribution enforces energy conservation. The function $\big|f_{1 \rightarrow 2}(\vec{q})\big|^{2}$ is the scalar atomic form factor \cite{Essig2016, Catena2020}, whose expression is:

\begin{equation}
    f_{1 \rightarrow 2}(\vec{q}) = \int \frac{d^{3}k}{(2\pi)^{3}}\psi_{2}^{*}(\vec{k} + \vec{q})\psi_{1}(\vec{k}).
    \label{eq:f_ion_definition}
\end{equation}
Here $1 \rightarrow 2$ denotes a transition from an initial state $1$ to a final state $2$. Before specializing to ionization, we relate this cross section to the event rate for an arbitrary electronic excitation:

\begin{equation}
    R_{1 \rightarrow 2} = \frac{\rho_{\chi}}{m_{\chi}}\left<\sigma v_{1 \rightarrow 2}\right>.
    \label{eq:R}
\end{equation}
The velocity average is performed with the dark matter distribution function. Substituting Eq. (\ref{eq:sigma_v}) into Eq. (\ref{eq:R}) yields:

\begin{equation}
    R_{1 \rightarrow 2} = \frac{\rho_{\chi}}{16\pi m_{\chi}^{3}m_{e}^{2}}\int d^{3}v\frac{d^{3}q}{4\pi}f_{\chi}(\vec{v})\delta\left(\Delta E_{1 \rightarrow 2} + \frac{q^{2}}{2m_{\chi}} - qv\cos(\theta_{qv})\right)\Big|M(E_{R}, v, \cos\theta)\Big|^{2}\Big|f_{1 \rightarrow 2}(\vec{q})\Big|^{2}.
    \label{eq:R_1_2}
\end{equation}

Following \cite{Essig2016}, the event rate depends on the detector orientation with respect to the Galactic rest frame \cite{Essig2012}. We therefore assume a spherically symmetric dark matter velocity distribution, which makes the calculation tractable. The delta function can be rewritten as:

\begin{equation}
    \delta\left(\Delta E_{1 \rightarrow 2} + \frac{q^{2}}{2m_{\chi}} - qv\cos(\theta_{qv})\right) = \frac{1}{vq}\delta\left(\cos(\theta_{qv}) - \cos(\theta_{qv}^{0})\right),
    \label{eq:change_of_delta}
\end{equation}
which allows the angular integration to be performed. Here:

\begin{equation}
    \cos(\theta_{qv}^{0}) = \frac{\Delta E_{1 \rightarrow 2}}{qv} + \frac{q}{2m_{\chi}v}.
    \label{eq:cos_theta_0}
\end{equation}
This expression defines the minimum lab-frame speed required to produce the transition $1 \rightarrow 2$ for given $q$ and $m_{\chi}$:

\begin{equation}
    v_{\mathrm{min}}(q) = \frac{\Delta E_{1 \rightarrow 2}}{q} + \frac{q}{2m_{\chi}},
    \label{eq:v_min}
\end{equation}
and for $\theta_{qv}^{0}=0$ it coincides with Eq. (\ref{eq:cos_theta_0}). Applying Eq. (\ref{eq:change_of_delta}) to Eq. (\ref{eq:R_1_2}) gives:

\begin{equation}
    R_{1 \rightarrow 2} = \frac{\rho_{\chi}}{16\pi m_{\chi}^{3}m_{e}^{2}}\int \frac{d^{3}q}{4\pi}dvd\theta_{qv}d\phi_{v}v^{2}\sin(\theta_{qv})f_{\chi}(v)\frac{1}{vq}\delta\left(\cos(\theta_{qv}) - \cos(\theta_{qv}^{0})\right)\Big|M(E_{R}, v, \cos\theta)\Big|^{2}\Big|f_{1 \rightarrow 2}(\vec{q})\Big|^{2}.
\end{equation}

Using $-d(\cos\theta_{qv})=\sin\theta_{qv}d\theta_{qv}$, the angular integration over $\cos\theta_{qv}$ is performed directly with the delta function. Regarding the threshold condition in Eq. (\ref{eq:v_min}), it is imposed with a step function, yielding:

\begin{align}
    R_{1 \rightarrow 2} &= \frac{\rho_{\chi}}{16\pi m_{\chi}^{3}m_{e}^{2}}\int \frac{d^{3}q}{4\pi}dvd\theta_{qv}d\phi_{v}v^{2}\sin(\theta_{qv})f_{\chi}(v)\frac{1}{2vq}\Theta(v - v_{\mathrm{min}}(q))\Big|M(E_{R}, v, \cos\theta)\Big|^{2}\Big|f_{1 \rightarrow 2}(\vec{q})\Big|^{2},\notag \\
    &= \frac{\rho_{\chi}}{32\pi m_{\chi}^{3}m_{e}^{2}}\int \frac{d^{3}q}{4\pi q}\eta(E_{R}, q)\Big|f_{1 \rightarrow 2}(\vec{q})\Big|^{2}.
\end{align}
Here we used:

\begin{equation}
    \int_{0}^{\pi} \sin(\theta_{qv}) d\theta_{qv} = 2,
\end{equation}
and defined:

\begin{equation}
    \eta(E_{R}, q) = \int \frac{d^{3}v}{v}f_{\chi}(v)\Big|M(E_{R}, v, \cos\theta)\Big|^{2}\Theta(v - v_{\mathrm{min}}(q)).
    \label{eq:eta}
\end{equation}
The squared matrix element can be expanded in powers of $v$ as in Eq. (\ref{eq:M^2}); applying that expansion to Eq. (\ref{eq:eta}) gives:

\begin{align}
    \eta(E_{R}, q) &= \int \frac{d^{3}v}{v}f_{\chi}(v)A\sum_{i = 0}^{10} B_{i}(E_{R}, \mathrm{cos}\ \theta)v^{i}\Theta(v - v_{\mathrm{min}}(q)), \\
    &\approx A B_{0}(E_{R}) \int \frac{d^{3}v}{v}f_{\chi}(v)\Theta(v - v_{\mathrm{min}}(q)) 
    \\
    & = A B_{0}(E_{R})\eta_{0}(v_{\mathrm{min}}(q)).\label{eq:eta_final}
\end{align}
As mentioned in the main text, terms of $\mathcal{O}(v)$ and higher were neglected. The remaining velocity dependence is then encoded in:

\begin{equation}
    \eta_{0}(v_{\mathrm{min}}(q)) = \int \frac{d^{3}v}{v}f_{\chi}(v)\Theta(v - v_{\mathrm{min}}(q)).
    \label{eq:eta_0}
\end{equation}

For ionization, $R_{1 \rightarrow 2}$ must be replaced by $R_{\mathrm{ion}}$, where the final state is a continuum positive-energy electron rather than a bound atomic level. The continuum electron is described by the wave function $\tilde{\psi}_{k'l'm'}(\vec{x})$ (see App. \ref{ap:radial_wave_functions}), with angular quantum numbers $(l',m')$ and asymptotic momentum magnitude $k'$. The ionization rate is obtained by summing over occupied initial states and integrating over all allowed final continuum states:

\begin{equation}
    R_{\mathrm{ion}}^{nl} = \sum_{s=-1/2}^{1/2}\sum_{m=-l}^{l}\sum_{l'=0}^{\infty}\sum_{m'=-l'}^{l'}\int dk'\frac{k'^{2}}{(2\pi)^{3}}R_{1 \rightarrow 2}.
\end{equation}
Formally, $1 \equiv (n,l)$ and $2 \equiv (k', l', m')$, so that:

\begin{equation}
    R_{\mathrm{ion}}^{nl} = \sum_{s=-1/2}^{1/2}\sum_{m=-l}^{l}\sum_{l'=0}^{\infty}\sum_{m'=-l'}^{l'}\int dk'\frac{k'^{2}}{(2\pi)^{3}}R_{n,l \rightarrow k'l'm'}.
\end{equation}

Now, changing variables from $k'$ to the natural logarithm of the recoil energy $\mathrm{ln}E_{R}$, we get:

\begin{equation}
    R_{\mathrm{ion}}^{nl} = \frac{\rho_{\chi}}{64\pi m_{\chi}^{3}m_{e}^{2}}\sum_{s=-1/2}^{1/2}\sum_{m=-l}^{l}\sum_{l'=0}^{\infty}\sum_{m'=-l'}^{l'}\int d\mathrm{ln}E_{R} \frac{d^{3}q}{4\pi q}\frac{k'^{3}}{(2\pi)^{3}}
    \eta(E_{R}, q)\Big|f_{n,l \rightarrow E_{R}l'm'}(\vec{q})\Big|^{2}.
    \label{eq:R_ion_before_f_ion}
\end{equation}
Given the summation involved in the equation above, it can be defined the dimensionless \textit{ionization form factor}, that is:

\begin{align}
    \Big|f_{\mathrm{ion}}(E_{R},q)\Big|^{2} &= \frac{2k'^{3}}{(2\pi)^{3}}\sum_{s=-1/2}^{1/2}\sum_{m=-l}^{l}\sum_{l'=0}^{\infty}\sum_{m'=-l'}^{l'}\Big|f_{n,l \rightarrow E_{R}l'm'}(\vec{q})\Big|^{2}, \notag \\
    &= \frac{4k'^{3}}{(2\pi)^{3}}\sum_{m=-l}^{l}\sum_{l'=0}^{\infty}\sum_{m'=-l'}^{l'}\left|\int d^{3}x \tilde{\psi}_{E_{R}l'm'}^{*}(\vec{x})\tilde{\psi}_{nlm}(\vec{x})\mathrm{e}^{i\vec{q}\cdot \vec{x}}\right|^{2}.
\end{align}
The spin sum is trivial because the wave functions are spin independent. The atomic scalar form factor is obtained by converting the Fourier representation to coordinate space. This quantity was computed in Sec. 3.a of App. B of \cite{Catena2020} and reads:

\begin{equation}
    \Big|f_{\mathrm{ion}}^{nl}(E_{R}, q)\Big|^{2} = \frac{4k'^{3}}{(2\pi)^{3}}\sum_{l'=0}^{\infty}\sum_{L=|l-l'|}^{l+l'}(2l + 1)(2l' + 1)(2L + 1) \begin{pmatrix}
        l & l' & L \\
        0 & 0 & 0
    \end{pmatrix}
    \big|I_{1}(q)\big|^{2}.
    \label{eq:f_ion_ap}
\end{equation}
As introduced in Sec. \ref{sec:physical_context_and_formulae}, $k' = \sqrt{2m_{e}E_{R}}$. The bracketed factor is the Wigner $3j$ symbol, and the overlap function $I_{1}$ depends on the radial wave functions:

\begin{equation}
    I_{1}(q) = \int dr r^{2}R_{k'l'}^{*}(r)R_{nl}(r)j_{L}(qr).
\end{equation}
Inside the integral above, the functions $j_{L}$ are the spherical Bessel functions. We have decided not to write explicitly all the dependencies in the name of the function $I_{1}$ for a matter of simplicity.

Since $I_{1}(q)$ depends only on the magnitude of $q$, the ionization form factor is independent of the direction of $\vec{q}$. Inserting Eq. (\ref{eq:f_ion_ap}) into Eq. (\ref{eq:R_ion_before_f_ion}) and performing the angular integration over $\vec{q}$ yields:

\begin{equation}
    \frac{dR_{\mathrm{ion}}^{nl}}{dE_{R}} = \frac{\rho_{\chi}}{128\pi m_{\chi}^{3}m_{e}^{2}}\int dq\ \frac{q}{E_{R}}\eta(E_{R}, q)\Big|f_{\mathrm{ion}}^{nl}(E_{R}, q)\Big|^{2}.
    \label{eq:dR_ion_Essig}
\end{equation}

The prediction of the quantity given in Eq. (\ref{eq:dR_ion_Essig}) has to be in the framework described in Sec. \ref{sec:physical_context_and_formulae}. In this case, the integral over the DM velocity is trivial since Dirac's delta representing the energy conservation does not depend on $\vec{v}$ at leading order according to the approximation of Eq. (\ref{eq:energy_conservation_approximation}). Therefore, it is valid to do:

\begin{equation}
    \int d^{3}v\ v^{n} f_{\chi}(v)\delta(m_{\chi} + E_{B}^{nl} - E_{R} - q) = \mu_{n}\delta(m_{\chi} + E_{B}^{nl} - E_{R} - q),
    \label{eq:int_1}
\end{equation}
where $\mu_{n}$ is the $n$-th moment of the distribution function $f_{\chi}(v)$. Repeating the calculation for the full energy-conservation delta function gives:

\begin{align}
    \int d^{3}v\ v^{n} f_{\chi}(v)\delta\left(\Delta E_{1 \rightarrow 2} + \frac{q^{2}}{2m_{\chi}} - qv\cos(\theta_{qv})\right) &= \int \frac{d^{3}v}{vq}v^{n} f_{\chi}(v)\delta\left(\cos(\theta_{qv}) - \cos(\theta_{qv}^{0})\right), \notag \\
    &= \frac{1}{2q}\int d^{3}v\ v^{n-1} f_{\chi}(v)\Theta(v - v_{\mathrm{min}}(E_{R}, q)), \notag \\
    &= \frac{\eta_{n}(v_{\mathrm{min}}(E_{R}, q))}{2q},
    \label{eq:int_2}
\end{align}
where $\eta_{n}$ was introduced in Eq. (\ref{eq:eta_0}) just for the case $n=0$. For the present approximation only this case is relevant, corresponding also to $n=0$ in Eq. (\ref{eq:int_1}). Therefore, under the assumption that Eq. (\ref{eq:energy_conservation_approximation}) holds, one may map:

\begin{equation}
    \eta_{0}(v_{\mathrm{min}}(E_{R}, q)) \rightarrow 2q\mu_{0} \delta(m_{\chi} + E_{B}^{nl} - E_{R} - q).\label{eq:maps}
\end{equation}

Using this mapping in Eq. (\ref{eq:eta_final}) yields:

\begin{equation}
    \eta(E_{R}, q) = 2qA B_{0}(E_{R})\mu_{0}\delta(m_{\chi} + E_{B}^{nl} - E_{R} - q).
\end{equation}
Since $\mu_{0}$ is the zeroth moment of a normalized distribution, it equals unity in this work.

It is also important to note that the factor $1/m_{\chi}^{2}m_{e}^{2}$ present in Eq. (\ref{eq:dR_ion_Essig}) must to be mapped to $1/m_{\chi}m_{e}^{2}q$ since the phase-space for fermionic absorption in \cite{Essig2016}, which is Lorentz invariant, differs from the case it is studied in this work. Therefore, applying this replacement to Eq. (\ref{eq:dR_ion_Essig}) gives:

\begin{align}
    \frac{dR_{\mathrm{ion}}^{nl}}{dE_{R}} &= \frac{\rho_{\chi}}{128\pi m_{\chi}^{2}m_{e}^{2}}\int dq\ \frac{\cancel{q}}{\cancel{q}E_{R}}2qA B_{0}(E_{R})\mu_{0}\delta(m_{\chi} + E_{B}^{nl} - E_{R} - q)\Big|f_{\mathrm{ion}}^{nl}(E_{R}, q)\Big|^{2},
    \\
    &= \frac{A\rho_{\chi}}{64\pi m_{\chi}^{2}m_{e}^{2}}B_{0}(E_{R})\mu_{0}\int dq\ \frac{q}{E_{R}} \delta(m_{\chi} + E_{B}^{nl} - E_{R} - q)\Big|f_{\mathrm{ion}}^{nl}(E_{R}, q)\Big|^{2}.
\end{align}
Integrating over $q$ using the delta function yields the following equation:

\begin{equation}
    \frac{dR_{\mathrm{ion}}^{nl}}{dE_{R}} = \frac{A\rho_{\chi}}{64\pi m_{\chi}^{2}m_{e}^{2}}B_{0}(E_{R})\mu_{0}\frac{q}{E_{R}}\Big|f_{\mathrm{ion}}^{nl}(E_{R}, q)\Big|^{2},
\end{equation}
where the right-hand side of the equation is evaluated in $q = m_{\chi} + E_{B}^{nl} - E_{R}$. This gives the differential ionization event rates in the framework of a massive DM neutral right-handed fermion interacting with electrons in the $(n, l)$ shell of a xenon atom, supposing that $m_{\chi}v \ll q$, and only as a function of the recoil energy $E_{R}$.

\section{Radial wave functions}\label{ap:radial_wave_functions}

\begin{figure*}
    \centering
    \includegraphics[width=0.495\textwidth]{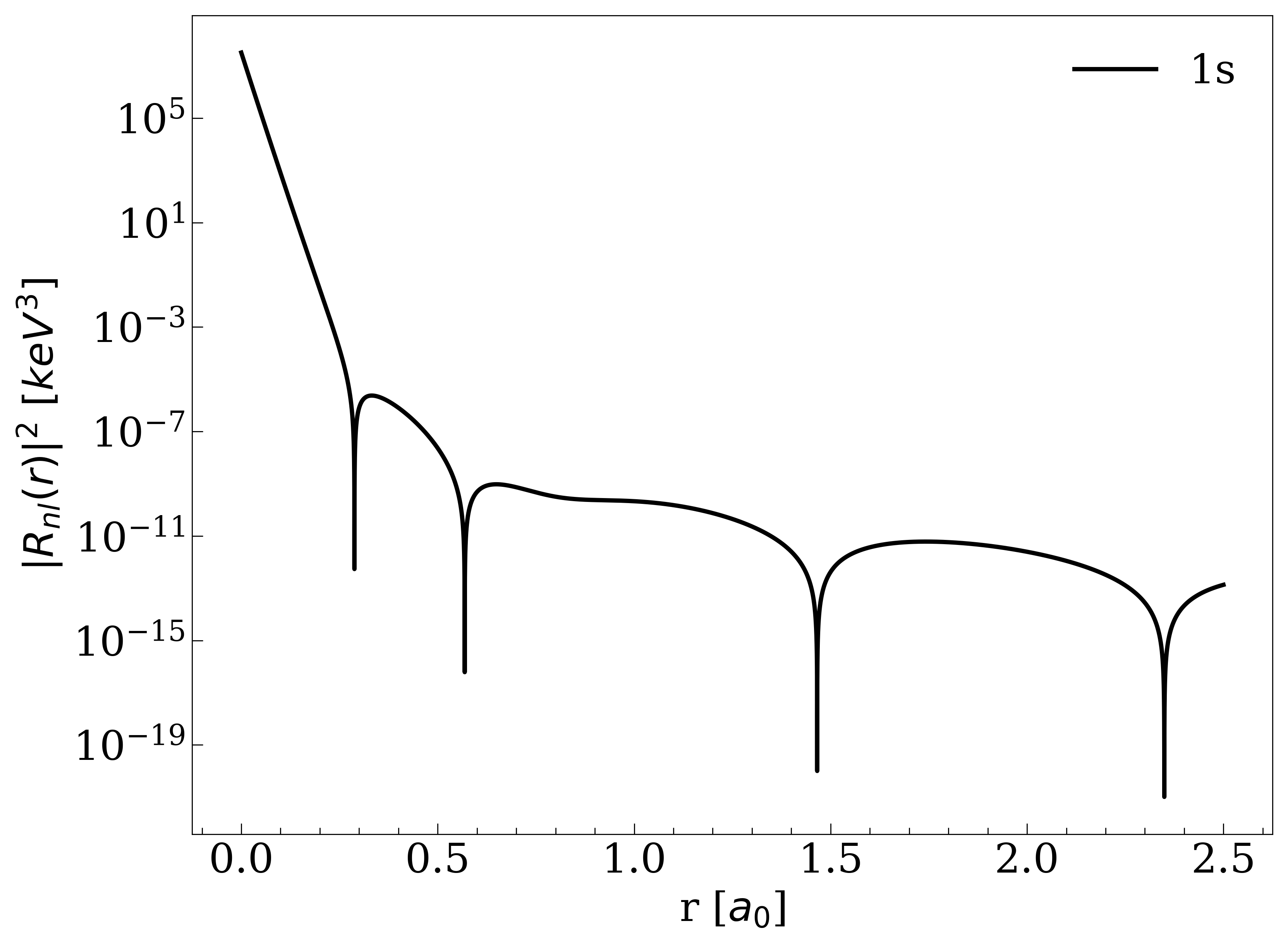}
    \hfill
    \includegraphics[width=0.495\textwidth]{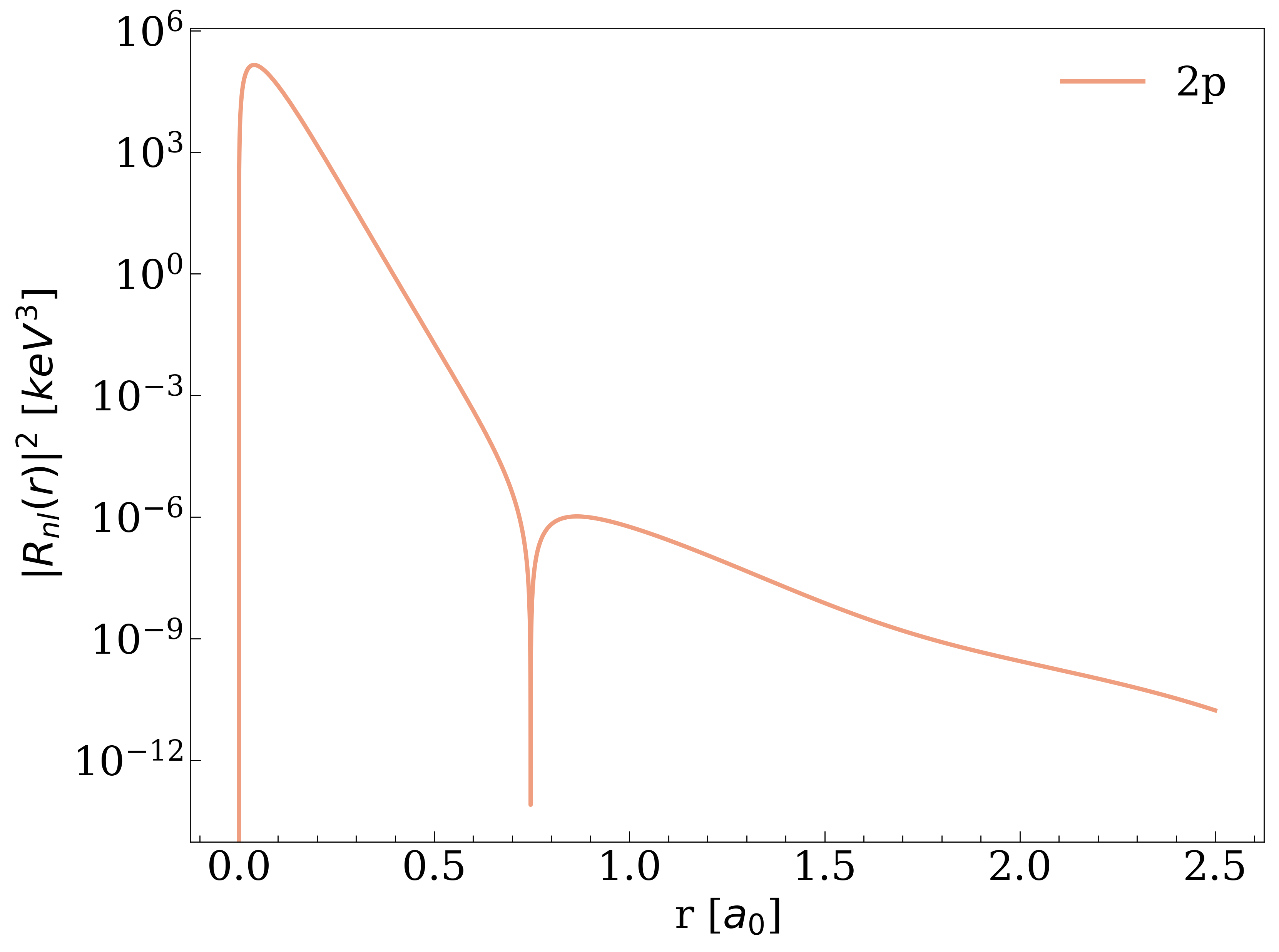}
    \vskip\baselineskip
    \includegraphics[width=0.495\textwidth]{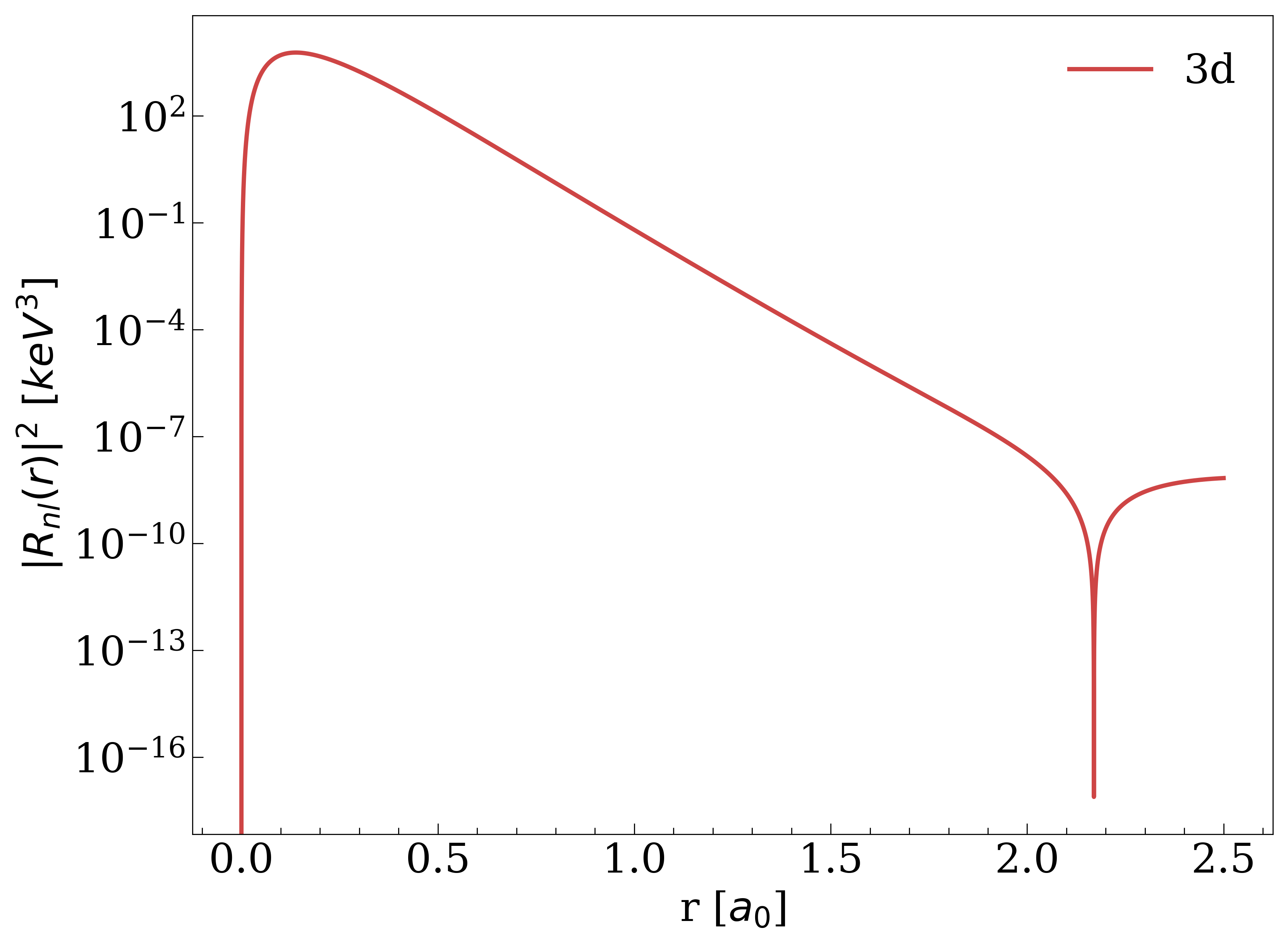}
    \hfill
    \includegraphics[width=0.495\textwidth]{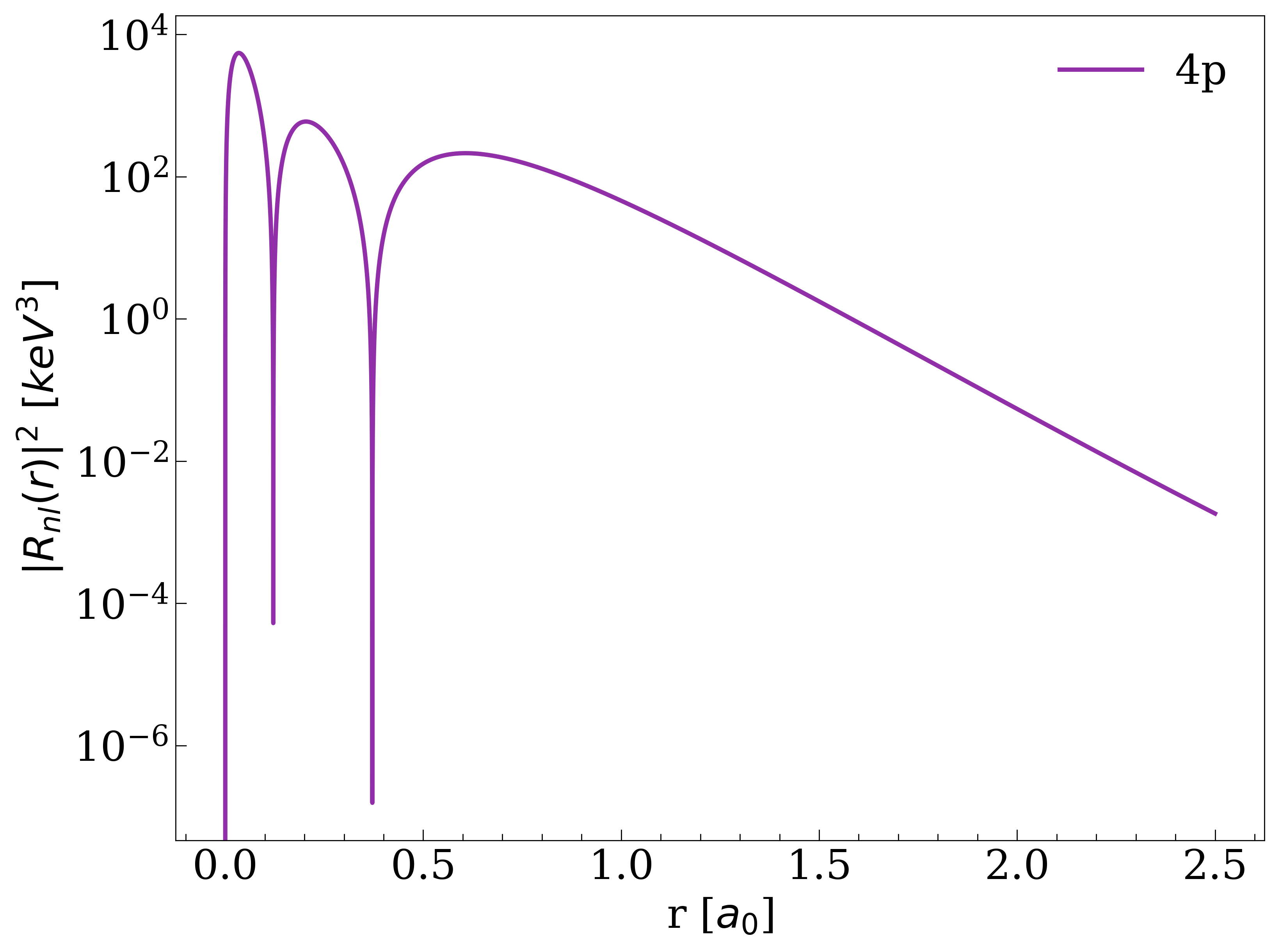}
    \caption{Radial wave functions of the bound electron, given for four different atomic shells. Note that these functions have physical units due to the Bohr radius.}
    \label{fig:r_in}
\end{figure*}

The ionization form factor introduced in App. \ref{ap:event_rates} requires both bound and continuum radial electron wave functions. The bound states are described by Roothaan-Hartree-Fock (RHF) radial functions expressed as linear combinations of Slater-type orbitals:

\begin{equation}
    R_{nl}(r) = a_{0}^{-3/2}\sum_{j}C_{jln}\frac{\left(2Z_{jl}\right)^{n_{jl}'+1/2}}{\sqrt{\left(2n_{jl}'\right)!}}\left(\frac{r}{a_{0}}\right)^{n_{jl}-1}.
\end{equation}
Here $a_{0}$ is the Bohr radius, and the coefficients $C_{jln}$, $Z_{jl}$, and $n_{jl}'$ are tabulated for xenon in \cite{Bunge1993}, as are the binding energies $E_{B}^{nl}$. RHF orbitals approximate Hartree-Fock wave functions using Slater determinants and thus respect the Pauli principle. The advantage of using the tabulated RHF coefficients in \cite{Bunge1993} relays in the energy errors of the functions being below $0.6$ meV. Four representative orbitals are shown in Fig. \ref{fig:r_in} for the $1s$, $2p$, $3d$, and $4p$ shells.

\begin{figure*}
    \centering
    \includegraphics[width=0.495\textwidth]{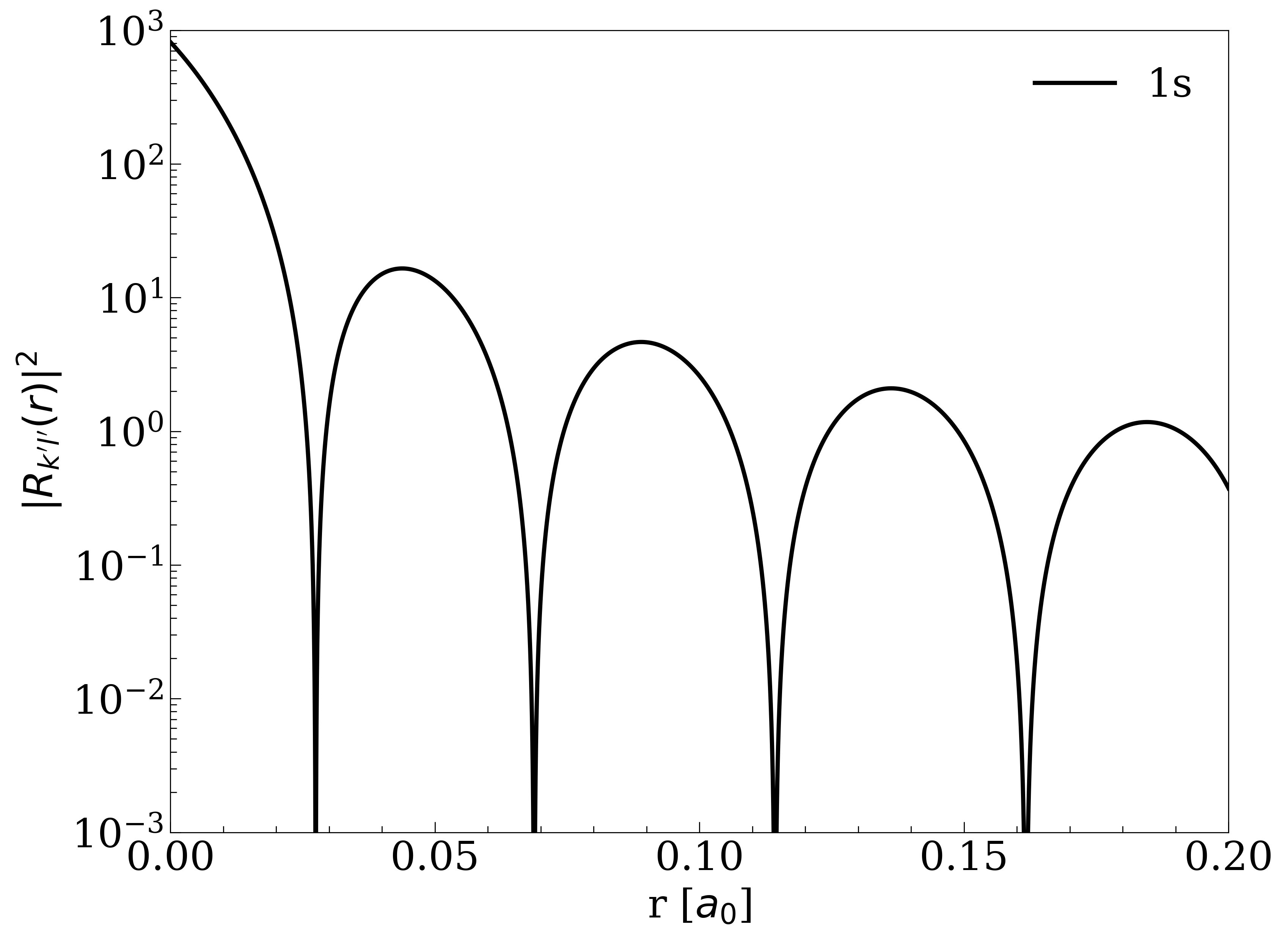}
    \hfill
    \includegraphics[width=0.495\textwidth]{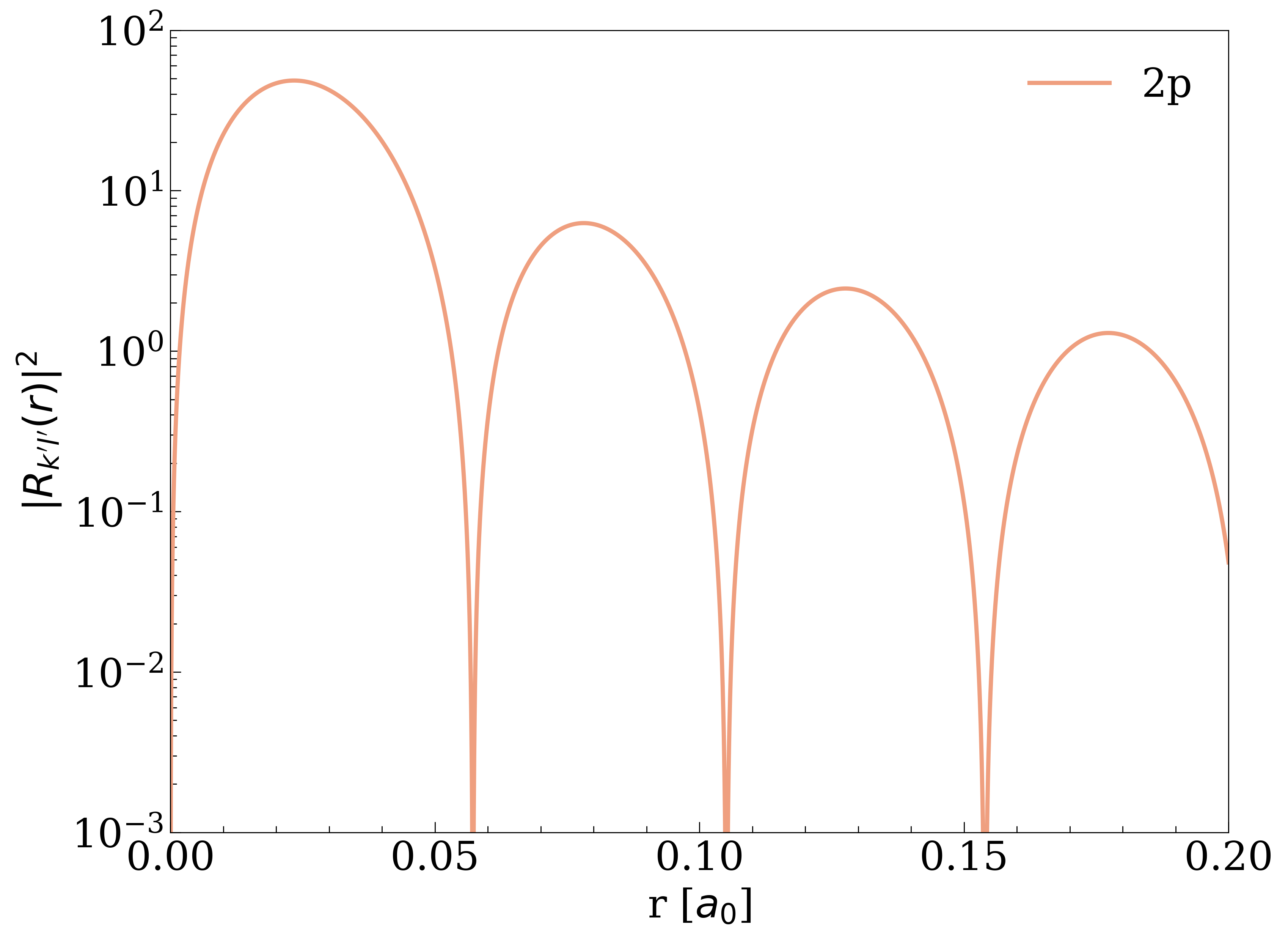}
    \vskip\baselineskip
    \includegraphics[width=0.495\textwidth]{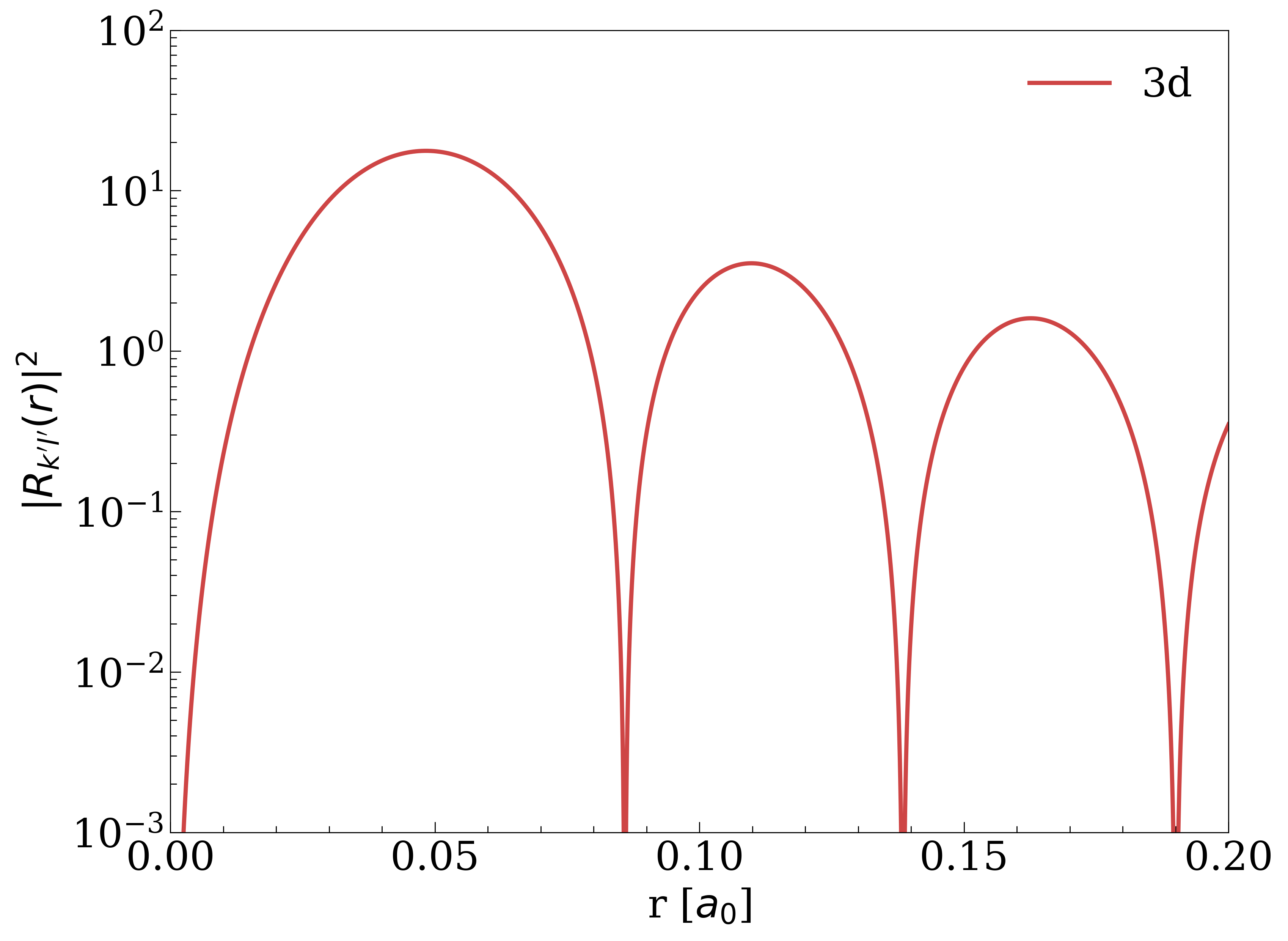}
    \hfill
    \includegraphics[width=0.495\textwidth]{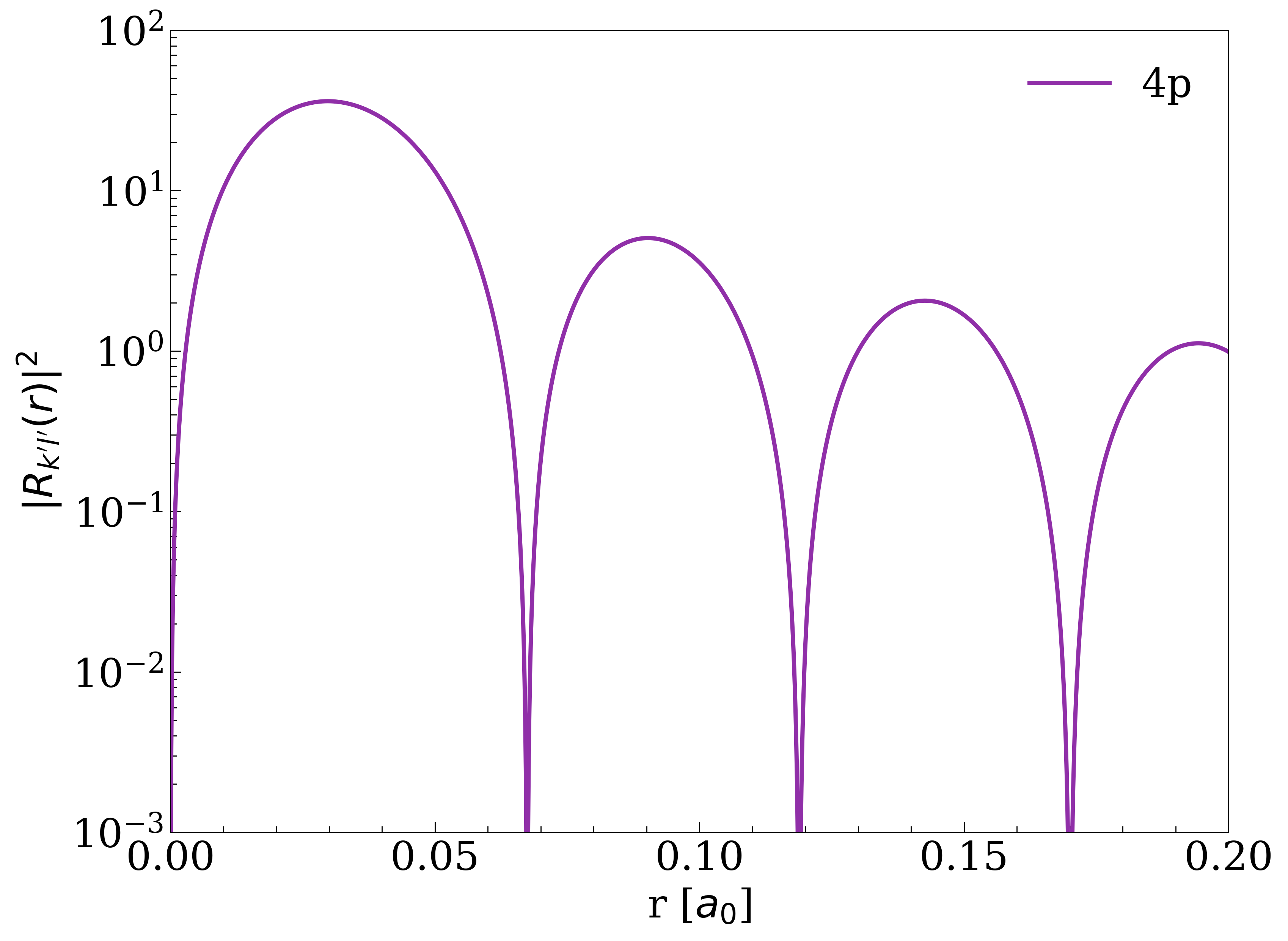}
    \caption{Outgoing radial wave functions computed for four different shells. The quantum numbers where fixed to the values $E_{R} = 50$ keV and $l' = l$ for all the cases.}
    \label{fig:r_out}
\end{figure*}

The unbound ionized electron is described by the Schrödinger equation with an effective central potential $Z_{\mathrm{eff}}(r)/r$. This gives positive-energy continuum states that account for the residual atomic potential. We approximate ionized xenon as hydrogen-like \cite{Li2022}, so the continuum electron sees a potential $-Z_{\mathrm{eff}}^{nl}/r$, where the effective charge $Z_{\mathrm{eff}}^{nl}$ is shell-dependent and determined from the binding energies \cite{TheDarkSideCollaboration2018}:

\begin{equation}
    Z_{\mathrm{eff}}^{nl} = \sqrt{\frac{|E_{B}^{nl}|}{13.6\ \mathrm{eV}}}\cdot n.
\end{equation}
Given the values above, it is possible to fully compute the atomic radial wave functions for positive energy continuum states in a hydrogen-like atom \cite{Bethe1957, Bethe1977}:

\begin{equation}
    R_{k'l'}(r) = (2\pi)^{3/2}\left(2k'r\right)^{l'}\frac{\sqrt{\frac{2}{\pi}}\left|\Gamma\left(l' + 1 - \frac{iZ_{\mathrm{eff}}}{k'a_{0}}\right)\right|}{(2l' + 1)!}\mathrm{e}^{\frac{\pi Z_{\mathrm{eff}}}{2k'a_{0}}-ik'r}{}_{1}F_{1}\left(l' + 1 + \frac{iZ_{\mathrm{eff}}}{k'a_{0}}, 2l' + 2, 2ik'r\right).
\end{equation}
The effective potential is not written in terms of the atomic numbers $(n, l)$ for the sake of clarity, but it does depend on the shell where the ionized electron was before the collision. The independent variable is $r$, the radial coordinate. The function ${}_{1}F_{1}(a, b, z)$ is the first kind confluent hypergeometric function. The quantum numbers $k'$ and $l'$ are, respectively, the momentum of the outgoing electron and its orbital quantum number. An illustration of the shape of these functions can be seen in Fig. \ref{fig:r_out}. There, the outgoing radial wave functions were computed for $E_{R} = 50$ keV and $l'= l$ for four different shells, $1s$, $2p$, $3d$, and $4p$. Recall that $k' = \sqrt{2m_{e}E_{R}}$.

\twocolumngrid
\bibliography{biblio}

\end{document}